\documentclass[prc,twocolumn]{revtex4}
\usepackage[pdftex]{graphicx}
\usepackage{amsfonts}
\usepackage{amsmath}
\usepackage{amssymb}
\usepackage{amsthm}
\usepackage{bm}

\newcommand{\tx}{\tilde{x}}
\newcommand{\tlt}{\tilde{t}}

\date{\today}
\begin{document}

\title{Photon HBT interferometry for non-central heavy-ion collisions}

\author{Evan Frodermann}
\email[Correspond to\ ]{frodermann@physics.umn.edu}
\affiliation{Department of Physics \& Astronomy, University of Minnesota,  
Minneapolis, MN 55455}
\affiliation{Department of Physics, The Ohio State University, 
Columbus, OH 43210}
\author{Ulrich Heinz}
\affiliation{Department of Physics, The Ohio State University, 
Columbus, OH 43210}

\begin{abstract}
Currently, the only known way to obtain experimental information about 
the space-time structure of a heavy-ion collision is through 2-particle 
momentum correlations. Azimuthally sensitive HBT interferometry 
(Hanbury Brown -- Twiss intensity interferometry) can complement 
elliptic flow measurements by constraining the spatial deformation
of the source and its time evolution. Performing these measurements on
photons allows us to access the fireball evolution at earlier times than
with hadrons. Using ideal hydrodynamics to model the space-time evolution 
of the collision fireball, we explore theoretically various aspects of 
2-photon intensity interferometry with transverse momenta up to 2 GeV,
in particular the azimuthal angle dependence of the HBT radii in 
non-central collisions. We highlight the dual nature of thermal photon 
emission, in both central and non-central collisions, resulting from the 
superposition of QGP and hadron resonance gas photon production. This 
signature is present in both the thermal photon source function and the 
HBT radii extracted from Gaussian fits of the 2-photon correlation function.
\end{abstract}


\maketitle

\section{Introduction}
Ideal hydrodynamic calculations \cite{Kolb:1999it,Kolb:2000sd,Teaney:2000cw,%
Teaney:2001av,Huovinen:2001cy,Hirano:2002ds,Kolb:2002ve} of heavy-ion 
collisions at the Relativistic Heavy Ion Collider (RHIC) generate momentum 
distributions of particles that correctly describe the effects of the 
collision fireball dynamics on the elliptic flow signature and successfully 
reproduce the main aspects of the measured momentum spectra for a large 
number of hadron species \cite{Ackermann:2000tr,Adler:2001nb,Adler:2001aq,%
Adcox:2001mf,Adams:2005dq,Adcox:2004mh,Huovinen:2003fa,Kolb:2003dz}. The 
good agreement between theory and experiment of the single-particle momentum 
distributions leaves, however, the fireball geometry essentially 
unconstrained. To eliminate the geometric ambiguities, spatio-temporal 
aspects of the reaction 
must be explored \cite{Lisa:2005dd}. The only known way to obtain experimental 
information on the space-time structure of the particle emitting source in
heavy-ion collisions is through two-particle momentum correlations 
\cite{Boal:1990yh,Heinz:1996rw,Heinz:1996bs}. The method of two-particle 
intensity interferometry, originally developed by Hanbury Brown and Twiss 
(HBT) \cite{Hanbury-Brown:1956pf} to measure angular distances of stars and 
other stellar objects, exploits quantum statistical correlations between 
particle intensities in two-particle coincidence measurements to access 
spatial information on the emitting source. Even though the space-time 
picture of the source extracted from intensity interferometry is necessarily
incomplete \cite{Lisa:2005dd,Heinz:1996bs,Heinz:1999rw,Wiedemann:1999qn}, it 
yields powerful geometric constraints which supplement the momentum-space 
information contained in the single particle spectra.

However, most theoretical calculations of the HBT radii, expressed through 
variances (Gaussian widths) of the source function, do not correctly
reproduce their measured momentum dependence \cite{Lisa:2005dd,Heinz:2002un}.
Various ways to explain this ``HBT puzzle'' and possibilities to correct 
this failure have been suggested. They include a microscopic treatment of 
the final freeze-out stage through hadronic rescattering \cite{Soff:2000eh}, 
exploration of fluctuations in the initial state \cite{Socolowski:2004hw}, 
different sets of initial conditions leading to stronger longitudinal 
\cite{Renk:2004yv} and transverse \cite{Broniowski:2008vp} hydrodynamic 
acceleration, as well as the inclusion of viscous effects 
\cite{Teaney:2003kp,Romatschke:2007jx,Pratt:2008sz,Pratt:2008qv} and final state 
interactions \cite{Cramer:2004ih,Pratt:2005hn,Kapusta:2005pt,Kuhlman:2007t}.
The influence on the characteristic HBT radii of finite resonance 
decay lengths after thermal freeze-out was explored \cite{Florkowski:2008cs,%
Kisiel:2008ws,Broniowski:2008vp}, and the impact of non-Gaussian features 
in the two-particle correlation function on the extraction of HBT radii 
from Gaussian fits to the correlator and their discrepancy from radii 
extracted through source function variances were studied 
\cite{Frodermann:2006sp}. Recent work \cite{Pratt:2008qv}
suggests that a comprehensive approach that includes all these effects
in a theoretically coherent fashion may be able to simultaneously 
describe single-hadron spectra and two-hadron correlations, lending 
support to the claim that hydrodynamic models successfully describe
both the dynamics and space-time structure (size, shape, and lifetime)
of the collision fireball at the point of hadron emission.

An interesting question that cannot be directly addressed by measuring
hadron distributions is the time-evolution of the collision fireball.
The reason is that hadrons interact strongly and decouple late. The 
observation of strong hadron elliptic flow \cite{Ackermann:2000tr}, i.e.  
a large momentum anisotropy in the final hadron spectra, has been linked
to early thermalization in the fireball and to almost perfect liquid 
behavior of the quark-gluon plasma (QGP) created in RHIC collisions
\cite{Heinz:2001xi,Gyulassy:2004zy}. This has opened the prospect
of exploring details of the QGP equation of state and of its conversion
to hadrons \cite{Kolb:2003dz} as well as its transport properties
(shear and bulk viscosity) through elliptic flow measurements 
\cite{Romatschke:2007mq,Song:2007fn,Luzum:2008cw,Song:2008hj}. Since the
QGP exists only in the early stage of the collision, a more direct probe 
of its properties and dynamical evolution is desirable rather than through 
soft hadron spectra emitted after its decay. We here discuss the emission
of thermal photons which occurs throughout the fireball evolution,
but most strongly during its hottest, earliest stage \cite{Shuryak:1978ij,%
Kajantie:1981wg,Kajantie:1986dh,Ruuskanen:1989tp,Ruuskanen:1993bk}.
Photons interact only electromagnetically and thus escape from the 
collision fireball without rescattering. For this advantage over hadrons
one has to pay with a correspondingly smaller production cross section which
makes direct photon measurements much more difficult than the observation
of soft hadrons.

The possibility of using photon elliptic flow measurements to access the
early evolution of momentum anisotropies in the expanding QGP fireball
was explored in \cite{Chatterjee:2005de,Heinz:2006qy,Chatterjee:2008tp}.
The growth of these momentum anisotropies is driven by spatial anisotropies 
of the pressure gradients in the initial hot source created in non-central
heavy-ion collisions. The collective flow response of the QGP to these
gradient anisotropies depends on its speed of sound and its viscosity
\cite{Song:2007ux,Song:2008si}. We propose that, in a similar way as
photon elliptic flow opens a window on the early evolution in momentum 
space, two-photon HBT interferometry may help to constrain the spatial
size and deformed shape of the fireball during this stage. By providing
access to both the space-time and momentum-space structure of the source 
at early times, thermal photons may thus help to constrain the validity
of hydrodynamic models, the QGP EOS and its transport properties during
the crucial hottest and densest period.

While photons are emitted throughout the evolution of the fireball and
thus do not allow for sharp snapshots of the early stage alone, their
emission rate depends on a high power of the fireball temperature
\cite{Shuryak:1978ij,Kajantie:1981wg,Kajantie:1986dh,Ruuskanen:1989tp,%
Ruuskanen:1993bk} which gives preference to early emission 
\cite{Srivastava:1993js,Srivastava:1993pt,Srivastava:1993hw,Srivastava:2004xp,%
Bass:2004de}. Indeed, we will show that the photon emission function has a 
distinctly bimodal structure,with two components that reflect emission of 
photon pairs of given pairmomentum from different spatial regions at early 
times (when radialcollective flow is still weak) and at late times (when the 
flow is strong).Although a single measurement is insufficient to cleanly 
separate these components, systematic exploration of the dependence of 
photon HBT radiion magnitude and direction of the photon pair momentum may 
eventually allow to do so.

The bimodal structure of the photon source is only one of two sources
for strong deviations of the emission function from a simple Gaussian
shape. The other is the massless nature of the photon. For two-photon
correlations it is therefore even more important than for hadrons that
the HBT radius parameters are computed through a procedure that matches
their experimental extraction. We here use a generalization to non-central
collisions of the Gaussian fit technique developed in 
\cite{Frodermann:2006sp,Lisa:2007zz}. Since photons, unlike charged 
hadrons, do not suffer from final-state Coulomb interactions, this 
technique is even more appropriate here than it is for hadron correlations.

Our study is exploratory in nature and, as such, lacks many features needed
for a quantitative analysis. To simulate the dynamical evolution of
the matter created in a heavy-ion collision we use {\tt AZHYDRO}, 
a (2+1)-dimensional code developed by P. Kolb \cite{Kolb:2000sd,Kolb:2002ve}
for solving the equations of {\it ideal} fluid dynamics in two transverse
dimensions assuming boost invariance along the longitudinal (beam) 
direction. We use the same photon emission rates as employed in earlier
studies of thermal photon spectra and elliptic flow \cite{Chatterjee:2005de,%
Turbide:2007mi,Gale:2008wf}. However, we exclude hard photons produced
before the fireball medium has thermalized in order to focus on thermal 
processes.. We also assume that photons
from the post-freeze-out decay of hadronic resonances are created 
sufficiently late that they do not correlate with thermal photons and
therefore do not affect the shape of the two-photon correlation function 
except at immeasurably small relative momenta. (They contribute to the 
single-photon spectra and will thus affect the normalization of the 
two-photon correlator, i.e. the correlation strength; this increases 
the statistics required for an accurate correlation measurement.)
Our equation of state assumes chemical equilibrium in the hadronic
phase below the critical temperature $T_c$ where the QGP converts
to hadrons, and thus it does not correctly reflect its measured
\cite{BraunMunzinger:2001ip} non-equilibrium chemical composition. All 
these deficiencies can be removed in time before two-photon correlation 
measurements become technically feasible.

\section{Two-particle correlators from non-central collisions} 
\label{GaussianFit}
\subsection{Correlation and emission functions}
\label{sec2a} 

If the emission function is a perfect Gaussian in space-time, the two-particle 
correlation function is a perfect Gaussian in relative momentum, and the HBT 
radii can be directly computed from the RMS variances of the emission 
function \cite{Wiedemann:1999qn}. Since, however, real emission functions
are seldom Gaussian, the direct comparison of RMS variances with the 
experimentally extracted HBT radii should in general be viewed with 
suspicion. A more reliable (even if more laborious) approach computes
from the emission function the actual correlation function and then
performs a Gaussian fit of the latter, using the same fit algorithm
as employed in experiment \cite{Frodermann:2006sp}. 

The two-particle correlation function for identical particles with momenta
$p_a$ and $p_b$ is defined as a ratio between the two-particle coincidence 
cross section and the product of the two single particle spectra as 
\begin{equation}
  C(\bm{p}_a,\bm{p}_b) = \frac{E_aE_b\frac{dN}{d^3p_ad^3p_b}}
                              {E_a\frac{dN}{d^3p_a}E_b\frac{dN}{d^3p_b}}\, .
\end{equation}
If the particles are emitted independently from the source, 
$C(\bm{p}_a,\bm{p}_b)$ can be calculated from the single-particle Wigner 
phase space density $S(x,K)$ (``emission function'') which describes the 
probability of emitting a particle from space-time point $x$ with momentum 
$K$, by folding it with the two-particle relative wave function 
\cite{Lisa:2005dd}. In the absence of final state interactions (as is
true for photon pairs) this wave function is a plane wave, yielding
\begin{eqnarray}
\label{eq:SPspectra}
  E\frac{dN}{d^3p} &=& \int d^4x\,S(x,p)\, ,
\\
\label{eq:GeneralCorrelator}
  C(\bm{q},\bm{K}) &=&\\ 
  &&\hspace{-1cm}1\pm \frac{1}{g_s}
  \frac{\left|\int d^4x\,S(x,K)\,e^{i\,q\cdot x}\right|^2}
       {\int d^4x\,S\left(x,K{+}\frac{q}{2}\right)
        \int d^4y\,S\left(y,K{-}\frac{q}{2}\right)},\nonumber
\end{eqnarray}
where $g_s$ is the spin degeneracy of the particles (for photons $g_s=2$),
and the $+$ ($-$) sign applies for bosons (fermions). From now on we will
only consider the bosonic case. 
 
The correlation function Eq.~\eqref{eq:GeneralCorrelator} depends on the 
relative momentum between the two particle $\bm{q}=\bm{p}_a{-}\bm{p}_b$, 
$q_0=E_a{-}E_b$ and the pair momentum $\bm{K}=(\bm{p}_a{+}\bm{p}_a)/2$, 
$K^0=(E_a{+}E_b)/2$. For the spectrum Eq.~\eqref{eq:SPspectra}, the 
emission function $S(x,K)$ must be evaluated on-shell ($K\mapsto p$), but 
for the correlation function Eq.~\eqref{eq:GeneralCorrelator} this is
not necessary \cite{Heinz:1996bs,Heinz:1999rw}. 

Since the measured momenta $p_a,\,p_b$ of the particles in the correlator 
are on-shell, the four-momenta $q$ and $K$ are necessarily off-shell. For 
pairs of identical particles the relative and pair momenta satisfy the 
orthogonality relation
\begin{equation}
  q^{\mu}K_{\mu}=0.
\label{eq:massshell}
\end{equation}
With this the argument of the plane wave becomes 
$q\cdot x={q^0}t-\bm{q}\cdot\bm{x}$, with 
$q^0={E}_a{-}E_b=\bm{\beta}\cdot\bm{q}$, where 
$\bm{\beta}=\bm{K}/K^0=2\bm{K}/(E_a{+}E_b)$. 

Two approximations are often used to further simplify 
Eq.~\eqref{eq:GeneralCorrelator}. The first is the ``smoothness 
approximation'' which assumes that the emission function varies slowly
over the momentum range where the correlator deviates from unity:
\begin{equation}
  C(\bm{q},\bm{K})\approx 1 + \frac{1}{g_s}
  \left|\frac{\int d^4x\,S(x,K)e^{i\,q\cdot x}}{\int d^4x\,S(x,K)}\right|^2.
\label{eq:TheoryDefinition}
\end{equation}
It is accurate as long as the curvature of the logarithm of the
single-particle spectrum is small \cite{Chapman:1994ax}. For thermal
sources created in heavy-ion collisions this is usually an excellent 
approximation \cite{Pratt:1997pw}. The second is the ``on-shell 
approximation'' $K^0\equiv(E_a{+}E_b)/2\approx E_K\equiv\sqrt{m^2+\bm{K}^2}$.
It allows us to replace the factor $\bm{\beta}$ in the relative wave function 
by the pair velocity and to substitute for the Wigner densities in 
Eq.~\eqref{eq:TheoryDefinition} the classical phase-space distributions 
for on-shell particles \cite{Heinz:2004qz}. Writing
\begin{eqnarray}
  K^0 &=& \frac{1}{2}(E_a+E_b)
\nonumber\\
\label{eq:kzero1}
      &\hspace{-0.7cm}=&\hspace{-0.5cm}\frac{E_K}{2}\left(
      \sqrt{1{+}\frac{q^2}{4\,E_K^2}{+}\frac{{\bm K}\cdot{\bm q}}{E_K^2}}
     +\sqrt{1{+}\frac{q^2}{4\,E_K^2}{-}\frac{{\bm K}\cdot{\bm q}}{E_K^2}} 
      \right)
\\\nonumber
      &\hspace{-0.7cm}=& \hspace{-0.5cm}E_K \left(1+\frac{q^2}{8\,E_K^2}\left(1{-}\cos^2\theta_{qK}\right)
          +\mathcal{O}\left(\frac{q^4}{E_K^4}\right)\right)
      \approx E_K\,,
\end{eqnarray}
we see that it applies as long as $q/(2E_K)\ll 1$, i.e. as long as the
source radius is much larger than the Compton wave length of the
particle pair \cite{Pratt:1997pw,Wiedemann:1999qn}. In heavy-ion 
collisions this holds for all hadron species (including pions) 
at all pair momenta $\bm{K}$, due to their large rest masses. For
massless photons, on the other hand, the on-shell approximation
breaks down at small pair momenta. For $m=0$ and $K\ll q/2$ one 
obtains instead of Eq.~\eqref{eq:kzero1}
\begin{eqnarray}
  K^0 &=& \\
\nonumber&&\hspace{-1cm}\frac{1}{2} \left(\frac{q}{2}\left(
      \sqrt{1{+}\frac{4K^2}{q^2}{+}\frac{4\bm{K}\cdot\bm{q}}{q^2}}
    + \sqrt{1{+}\frac{4K^2}{q^2}{-}\frac{4{\bm K\cdot\bm q}}{q^2}}
      \right)\right)
\nonumber\\
\label{eq:secondTaylor}
     &\hspace{-0.6cm}=& \hspace{-0.3cm}\frac{q}{2}
      \left[1+\frac{2K^2}{q^2}\bigl(1{-}\cos^2\theta_{qK}\bigr)
             +{\mathcal O}\left(\frac{K^4}{q^4}\right)\right]
      \approx \frac{q}{2}.
\end{eqnarray}
Hence, for small-momentum photon pairs $\bm{\beta} = \bm{K}/K^0\approx 
2\bm{K}/q$ (which obviously differs from the pair velocity which has
magnitude $c=1$). We will see that this has interesting consequences for 
the structure of the two-photon correlation function and the $K$-dependence 
of photon HBT radii.

\subsection{Gaussian sources and RMS variances}
\label{Sec:RMStheory}
%
The main characteristic of the correlation function is the $q$-range
over which it decays to 1. The corresponding half-widths of the correlator
are called ``HBT radius parameters'' or ``HBT radii''. For a Gaussian source
they can be related to the ``homogeneity lengths'' (defined below) of the 
source \cite{Heinz:1996bs}. In the absence of final state interactions
they can be extracted by fitting the correlator $C(\bm{q},\bm{K})$
with a Gaussian function of the form 
\begin{equation}
\label{eq:Gaussfit}
  C(\bm{q},\bm{K}) = 1 + \lambda(\bm{K})\,
  \exp{\left[-\sum_{ij=x,y,z}q_i\,q_j\,R_{ij}^2(\bm{K})\right]}\, ,
\end{equation}
where the correlation strength parameter $\lambda(\bm{K})$ is introduced 
to account for uncorrelated pairs that contribute to the single-particle
spectra in the denominator but not to the correlated part of the 
two-particle cross section in the numerator. Uncorrelated pairs arise
from far-separated decay products of long-lived resonances and (if the
particles carry spin) from pairs with unaligned particle spins (e.g.
photons with opposite helicity). The correlation strength is also 
reduced if particles are not emitted independently, but partially 
coherently \cite{Wiedemann:1999qn}.

In general the HBT radii $R^2_{ij}$ depend on the pair momentum. They are
interpreted as width parameters of the effective source (``homogeneity 
regions'' \cite{Makhlin:1987gm}) from which particles with momentum 
$\bm{K}$ are emitted. In collectively expanding sources, where the 
particle momenta are correlated with position by a boost with the local 
flow velocity, the homogeneity regions for particles with a given momentum
constitute only a fraction of the entire fireball. However, only the 
homogeneity lengths can be measured with HBT correlations. Furthermore, the HBT
radii measure only certain combinations of spatial and temporal width
parameters (``variances'') of the source, as imposed by the mass-shell 
constraint \eqref{eq:massshell} which implies $q\cdot x = 
-\bm{q}\cdot(\bm{x}{-}\bm{\beta}t)$:
\begin{eqnarray}
  R_{ij}^2(\bm{K})&=&\langle(\tx_i-\beta_i\tlt )(\tx_j-\beta_j\tlt)\rangle\nonumber\\
  &\equiv& -\frac12 \left.\frac{\partial^2 C(\bm{q},\bm{K})}
                             {\partial q_i \partial q_j}\right|_{\bm{q}{=}0}. 
\label{eq:derivativeRMS}
\end{eqnarray}
Here $\langle f \rangle$ denotes an average over the the emission 
function, 
\begin{equation}
\label{average}
  \langle f \rangle \equiv 
  \frac{\int d^4x\, f(x)\,S(x,\bm{K})}{\int d^4x\, S(x,\bm{K})}\,,
\end{equation}
and $\tilde{x}^\mu\equiv x^\mu-\langle x^\mu\rangle$. The relations  
\eqref{eq:derivativeRMS} are exact for Gaussian sources where the 
inverse width of the correlator agrees with its curvature at $\bm{q}=0$. 
In this case they can be used as a 'shortcut' for the calculation of HBT 
radii, by evaluating the RMS variances in \eqref{eq:derivativeRMS} directly 
from the emission function $S(x,K)$ instead of first computing the correlator 
from \eqref{eq:TheoryDefinition} and then fitting it with a Gaussian as 
in \eqref{eq:Gaussfit}. Here we will not use this shortcut but show 
results obtained with the help of Eq.~\eqref{eq:derivativeRMS} only 
for comparison purposes, to illustrate the importance of non-Gaussian 
features in the source and correlators.

\subsection{Gaussian fitting procedure for non-central collisions}
\label{sec2c}

The emission function $S(x,K)$ is computed from the hydrodynamic model 
{\tt AZHYDRO}, for hadrons as described in \cite{Kolb:2003dz} and for
photons as outlined in \cite{Chatterjee:2005de,Heinz:2006qy,%
Turbide:2007mi} which folds the thermal photon emission rate with the
hydrodynamic temperature and flow evolution. Using this emission
function in the average \eqref{average}, we compute the correlation 
function Eq.~\eqref{eq:TheoryDefinition} as
\begin{equation}
\label{eq:TheoryIdOnly}
g_s\bigl(C(\bm{q},\bm{K}) - 1\bigr) 
  = \langle\cos(q\cdot x)\rangle^2 + \langle\sin(q\cdot x)\rangle^2\\
\end{equation}
where
\begin{equation}
q\cdot x = (E_a{-}E_b)\,t - q_x\,x - q_y\,y - q_l\,z
\end{equation}
with $E_{a,b}^2 = E_K^2 \pm {\bm K}\cdot{\bm q}+{\bm q}^2/4$.
We will only consider ``mid-rapidity pairs'' with zero pair momentum 
along the beam direction ($K_l=0$) such that ${\bm K}\cdot{\bm q} = 
K_{\perp}\,q_{o}$ where $q_o$ denotes the ``outward'' component of the 
relative momentum (along the emission direction of the pair in the 
transverse plane). The ``sideward'' component $q_s$ is defined as the
one perpendicular to $\bm{K}_\perp$ and the beam direction.

For collisions between equal-mass nuclei and pairs emitted with zero
longitudinal momentum in the center-of-mass frame, the source is
reflection symmetric in the longitudinal direction, and the general 
form \eqref{eq:Gaussfit} reduces to \cite{Wiedemann:1997cr,Lisa:2000ip}
\begin{equation}
\label{eq:3dGaussNC}
  C(\bm{q},\bm{K}_\perp) = 
  1 + \lambda \, 
  e^{-\left(q_o^2R_o^2+q_s^2R_s^2
           +q_l^2R_l^2+2q_{o}q_{s}R_{os}^2
      \right)}\, ,
\end{equation}
where $\lambda$, $R_o$, $R_s$, $R_l$ and $R_{os}$ are all dependent on 
$\bm{K}_\perp$.

For central collisions between spherical nuclei, the emission function 
is azimuthally symmetric and all directions of $\bm{K}_\perp$ are 
equivalent (i.e. the correlator and HBT radii only depend on the magnitude
of $K_\perp$). As a consequence, the cross-term $R_{os}^2$ vanishes
\cite{Wiedemann:1997cr,Lisa:2000ip}. In non-central collisions, the 
source is deformed and initially out-of-plane elongated with respect to 
the reaction plane spanned by the impact parameter $\bm{b}$ (defining 
the $x$-axis) and the beam (defining the $z$-axis). As a result the 
cross-term $R_{os}^2$ is now non-zero, and the 2-particle correlation 
function \eqref{eq:TheoryIdOnly} and its HBT radii \eqref{eq:3dGaussNC}
now depend on the azimuthal angle $\Phi$ of the emission direction 
$\bm{K}_\perp$.  This dependence is both implicit through azimuthal symmetry 
violations in the emission function $S(x,K)$ with which the averages in 
\eqref{eq:TheoryIdOnly} are taken and explicit through the azimuthal 
rotation between the reaction-plane coordinate system (in which the 
source has been computed) and the $osl$ system defined by the emission 
direction of the pair in which the HBT radii \eqref{eq:3dGaussNC} are 
determined \cite{Wiedemann:1997cr,Lisa:2000ip,Heinz:2002au}:
\begin{eqnarray}
  q_x &=& q_{o}\,\cos\Phi - q_{s}\,\sin\Phi\,,
\nonumber\\
  q_y &=& q_{o}\,\sin\Phi + q_{s}\,\sin\Phi\,.
\label{eq:qxqydefs}
\end{eqnarray}

The HBT radii are determined by fitting the computed correlation
function with the functional form \eqref{eq:3dGaussNC}. To this end
we generalize the Gaussian fitting algorithm developed in 
\cite{Frodermann:2006sp,Lisa:2007zz} to include the $R_{os}$ as an 
additional fit parameter. (Note that $R_{os}$ cannot be determined from 
1-dimensional Gaussian fits to slices of the correlation function along
any one of the $osl$ axes; it requires at least a 2-dimensional Gaussian 
fit in the $os$ plane.) We write
\begin{eqnarray}
  \ln\left[g_s\bigl(C({\bm q})-1\bigr)\right] &=& \\
  &&\hspace{-2cm}\ln \lambda - (q_o^2R_o^2+q_s^2R_s^2+q_l^2R_l^2+2q_oq_sR_{os}^2),\nonumber
\end{eqnarray}
evaluate this expression on a suitable set of $\bm{q}$-points $\bm{q}^{(k)}$
($k=1,\dots,N$), and define $\chi^2$ as 
\begin{equation}
\label{eq:chisqNC}
  \chi^2 = \sum_{k=1}^N \left[\frac{\ln{[C(\bm{q}^{(k)})-1]}
  -\ln{\lambda}+M(\bm R,\bm{q}^{(k)})}{\sigma'_k}\right]^2\, ,
\end{equation}
with 
\begin{equation}
  M(\bm R,\bm{q}^{(k)}) = 2\,q_o^{(k)}q_s^{(k)}R_{os}^2
  +{\displaystyle{\sum_{i=o,s,l}}}\bigl(q_i^{(k)}\bigr)^2\,R_i^2\, .
\end{equation}
For the bin error $\sigma'_k$ we use \cite{Frodermann:2006sp}
\begin{equation}
\label{eq:sigma_prime}
  \sigma'_k = \frac{\sigma_k}{C(\bm{q}^{(k)})-1}\,.
\end{equation}
Minimization of $\chi^2$ with respect to the fit parameters,  
\begin{equation}
\label{eq:PartialsNC}
  \frac{\partial \chi^2}{\partial \ln{\lambda}} = 0\,, \qquad
  \frac{\partial \chi^2}{\partial R_i^2} = 0 \quad (i=o,s,l,os)\,, 
\end{equation}
produces a set of five coupled linear equations that can be written
in matrix form as
\begin{equation}
\label{eq:LinearEqs}
   \sum_{\beta} T_{\alpha\beta} P_{\beta} = V_{\alpha} \,,
\end{equation}
where $\alpha$ and $\beta$ take the values \o$,o,s,l, os$ (\o\ is associated
with the correlations strength $\lambda$). The vectors on the right
hand side have the form
\begin{eqnarray}
\label{eq:PNC}
  P &=& \left( \ln{\lambda} , R_o^2 , R_s^2 , R_l^2, R_{os}^2\right) ,
\\
  V_{\textrm{\o}} &=& -\sum_{k=1}^N\frac{\ln{[C(\bm{q}^{(k)})-1]}}
                                        {(\sigma'_k)^2} \,,
\\
\label{eq:VNC}
  V_i & = & + \sum_{k=1}^N\frac{\bigl(q_i^{(k)}\bigr)^2}{(\sigma'_k)^2}
  \cdot\ln{[C(\bm{q}^{(k)})-1]} ,
\\
  V_{os} & = & + 2\sum_{k=1}^N\frac{q_o^{(k)}q_s^{(k)}}
  {(\sigma'_k)^2}\cdot\ln{[C(\bm{q}^{(k)})-1]} ,
\end{eqnarray}
while the symmetric $5\times 5$ matrix $T$ has the components 
\begin{eqnarray}
\label{eq:TNC}
  T_{\textrm{\o}\textrm{\o}}  &=& - \sum_{k=1}^N\frac{1}{(\sigma'_k)^2}\,,
\nonumber \\
  T_{\textrm{\o}i}  &=& + \sum_{k=1}^N\frac{\bigl(q_i^{(k)}\bigr)^2}
                                           {(\sigma'_k)^2}\,,
\nonumber \\
  T_{ij} &=& - \sum_{k=1}^N
  \frac{\bigl(q_i^{(k)}\bigr)^2\,\bigl(q_j^{(k)}\bigr)^2}
       {(\sigma'_k)^2}\,,
\\
  T_{\textrm{\o},os}  &=& + 2\sum_{k=1}^N
  \frac{q_o^{(k)}q_s^{(k)}}{(\sigma'_k)^2}\,,
\nonumber \\
  T_{i,os} &=& - 2\sum_{k=1}^N
  \frac{\bigl(q_i^{(k)}\bigr)^2\,\bigl(q_o^{(k)}q_s^{(k)}\bigr)}
       {(\sigma'_k)^2}\,,
\nonumber
\end{eqnarray}
with $\{i,j\}$ = $(o,s,l)$. As before \cite{Frodermann:2006sp}, this set 
of linear equations is easily solved through matrix diagonalization of 
$T_{\alpha\beta}$.

\section{Two-pion correlations in non-central Au+Au collisions}
\label{Pions}

As a reference for comparison with the two-photon correlation functions
computed further below, we briefly review the behavior of two-pion
correlations and pion HBT radii in non-central Au+Au collisions at 
RHIC energies. This analysis complements earlier work \cite{Heinz:2002sq}
which was based on the same hydrodynamical model as used here but made use
of the shortcut (\ref{eq:derivativeRMS}) to calculate the HBT radii directly 
from the RMS variances of the hydrodynamic pion emission function. As 
explained above and numerically studied in \cite{Frodermann:2006sp}, this 
shortcut becomes doubtful when the emission function is not well described 
by a Gaussian in space-time. Here, we first compute the two-pion correlator 
(\ref{eq:TheoryDefinition}) numerically from the hydrodynamic emission 
function $S_\pi(x,K)$ \cite{Chapman:1994xa} and then obtain the HBT radii 
from a 3-d Gaussian fit to this correlation function as described in the 
preceding Section. For simplicity we include only directly emitted pions, 
i.e. we neglect pions from post-freeze-out decays of unstable resonances.  

\begin{figure*}
\includegraphics[width=\linewidth]{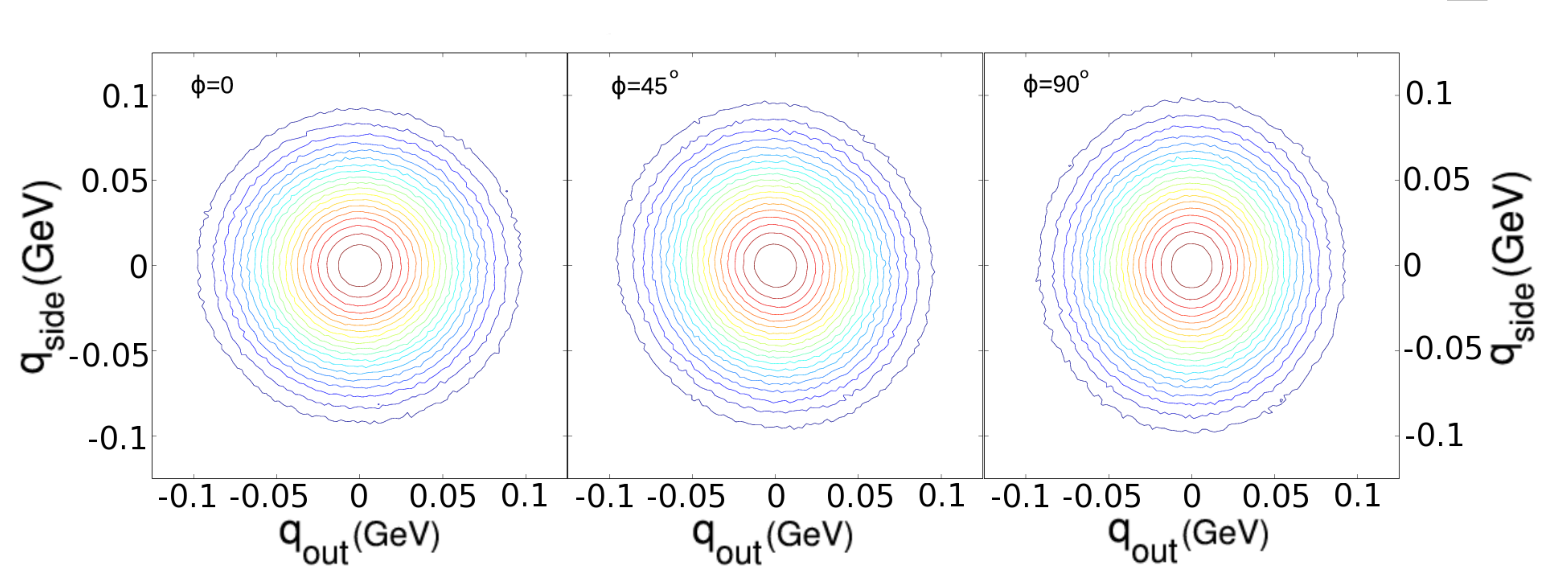}
\caption[Contour plots of the pion correlation function, C(qout, qside)-1]%
{(Color online)
Contour plots of the pion correlation function $C(q_o,q_s,q_l{=}0)-1$, at 
$\Phi=0^\circ,\,45^\circ$, and $90^\circ$ in the limit $K_\perp \to 0$, for 
200\,$A$\,GeV Au+Au collisions at $b=7$\,fm.
When rotating $\bm{K}_\perp$ counter-clockwise, the correlation function 
rotates clockwise. Successive contours are separated by 0.05, starting 
from $C(0)-1{\,=\,}1$ in the center.
\label{fg:CorrelPionNC}
}
\end{figure*}

Figure~\ref{fg:CorrelPionNC} shows the azimuthal angle dependence of the 
pion correlation function for peripheral Au+Au collisions at $b=7$\,fm, 
in the limit $K_\perp\rightarrow0$ as we change the direction of 
${\bm K}_\perp$. At $K_\perp=0$ the correlation function ``sees'' the 
entire fireball \cite{Heinz:2002sq}. Correspondingly, for non-central 
collisions at RHIC energies, the outward radius increases and the sideward 
radius decreases as we move away from $\Phi=0$, due to the out-of-plane 
deformation of the source at freeze-out. Beyond $\Phi=90^\circ$ these
tendencies reverse. Through geometric arguments one easily sees that these 
oscillations are seen as a clockwise rotation of the correlator contours, 
opposite to the counter-clockwise rotation of the direction of the 
transverse pair momentum. These same oscillations were observed in 
previous calculations that used RMS variances of the emission function
as proxies for the HBT radii \cite{Heinz:2002sq}. 

\begin{figure}
\begin{center}
\includegraphics[width=\linewidth]{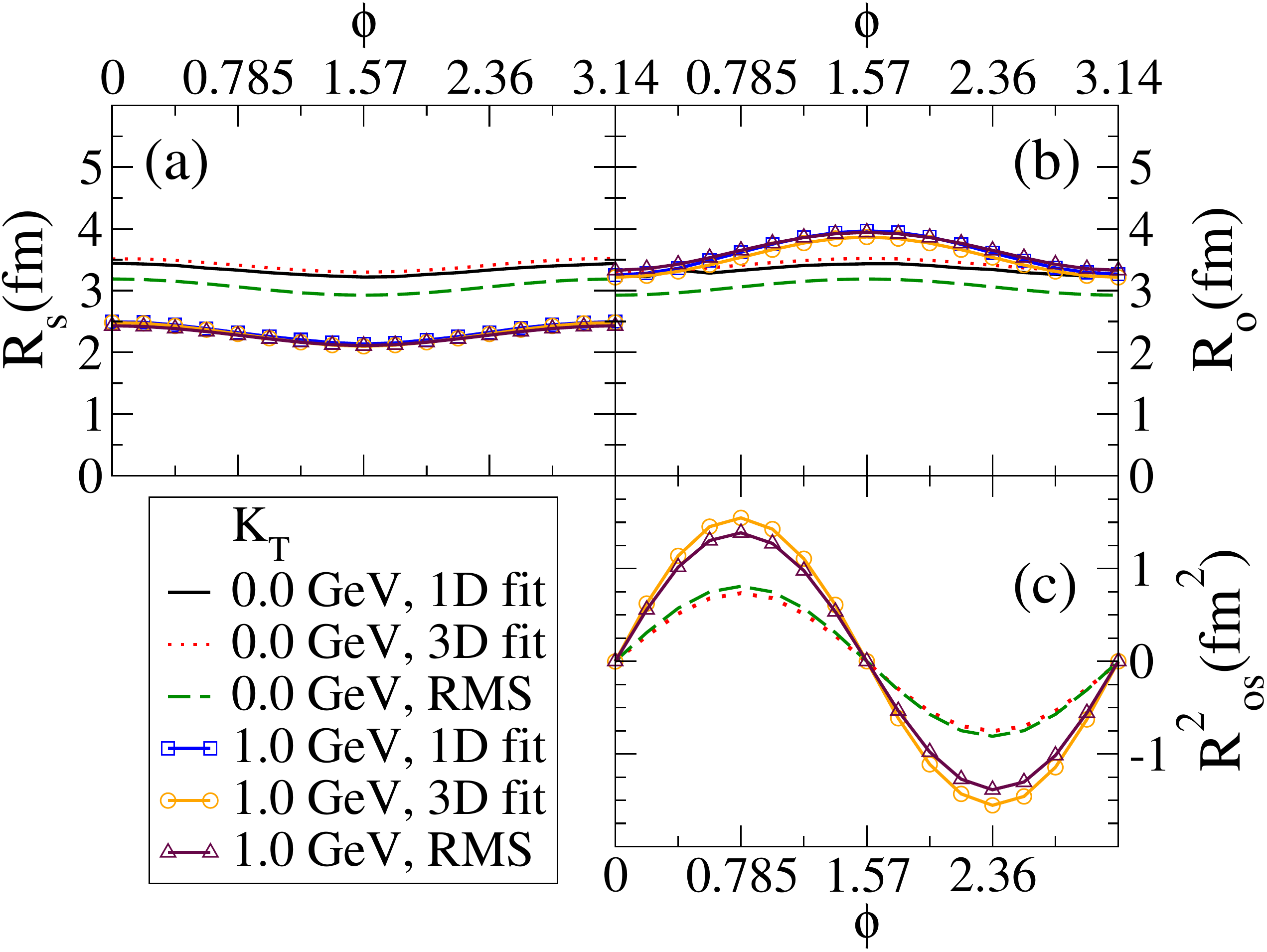}
\end{center}
\caption[Pion HBT oscillations for both fitted and RMS radii]{(Color online)
Pion HBT radii $R_o$, $R_s$, and $R_{os}^2$ from semi-peripheral ($b=7$\,fm ) 
Au+Au oscillations at $\sqrt{s_{NN}} = 200$\,GeV, as functions of the 
azimuthal emission angle $\Phi$. See text for discussion of different 
extraction methods.}
\label{fg:pionHBTNC}
\end{figure}

Figure~\ref{fg:pionHBTNC} compares the azimuthal oscillations of the pion 
HBT radii for 200\,$A$\,GeV Au+Au collisions at $b=7$\,fm with those of the 
RMS radii, for two values of $K_\perp$ (0 and 1\,GeV/$c$). In
addition to the results from the 3-dimensional Gaussian fit described in the 
previous Section (labeled ``3D'') we also show, for the purpose of 
comparison, radius parameters extracted from 1-dimensional Gaussian fits 
to slices of the correlation function along the $q_s$ and $q_o$ axes 
(labeled ``1D''). (Note that the $R^2_{os}$ cross term cannot be 
extracted from these slices, so the corresponding 1D curves are missing.) 
In the 3D Gaussian fits, we restrict the $q_l$ fit range to a narrow 
window $q_l < q_\mathrm{max} = 0.02$ GeV, in order to avoid distortions 
in $R_o$ arising from strong non-Gaussian tails of the correlator at 
larger $q_l$ values (see Ref.~\cite{Frodermann:2006sp} and 
Figure~\ref{fg:photonslice} below).
This is permissible since we found that the shape of the correlation 
function along the $q_l$ direction is almost independent of $\Phi$,
including its non-Gaussian tail. The reason is that, even though 
non-Gaussian effects of the source are strongest along the beam direction,
they are mainly caused by the strong boost-invariant longitudinal expansion 
of the fireball which is independent of impact parameter and transverse 
position in the reaction zone, and thus $\Phi$-independent. If we use a 
larger fit range in $q_l$, thereby capturing more of the non-Gaussian 
tail, we find an intercept parameter $\lambda<1$ that oscillates with 
$\Phi$. These oscillations interfere with the oscillations of the 
transverse radius parameters $R_o$ and $R_s$, significantly modifying 
their oscillation amplitudes. This is an undesirable effect caused by 
non-Gaussian features which lie entirely in the longitudinal domain. By 
restricting the fit range in the $q_l$ direction as described we can keep 
the correlation strength parameter close to 1 for all angles and ensure 
that the azimuthal oscillations of the transverse HBT radii faithfully 
reflect the oscillations in the widths of $C(q_o,q_s)$ as seen in the 
contour plots.

Figure~\ref{fg:pionHBTNC} shows that the {\em azimuthal oscillation 
amplitudes} of the HBT radii from the Gaussian fit agree very well with 
those of the RMS radii even though (as previously observed 
\cite{Frodermann:2006sp}) the {\em $\Phi$-averaged} HBT radii differ 
somewhat from the corresponding RMS radii. $R_{os}^2$ is seen to 
increase with $K_\perp$, continuing the previously predicted 
\cite{Heinz:2002sq,Frodermann:2007ab} and experimentally confirmed 
\cite{Adams:2003ra} trend to larger values of $K_T$. 

\section{Photon interferometry}
\label{PhotonTheory}

Since photons are emitted throughout the fireball evolution, with emission 
at higher transverse momenta weighted towards earlier times and higher
temperatures, we expect (and our calculations confirm) that, for 
sufficiently high $K_\perp$, the thermal photon HBT radii will reflect 
geometric and dynamical characteristics of the smaller and more deformed 
early source. This is a simple consequence of the dominance of QGP 
radiation in the single photon yield at high $p_\perp$. However, similar 
to pion interferometry, at any given $K_\perp$ contributions from lower 
temperature regions that are blue shifted by collective flow will mix with 
the early emission contributions. This lower temperature emission is 
generated at later times when the source is larger and less deformed. 
There are two competing processes; small radii with weak flow and 
large source deformations reflected in the early emitted photons and larger 
radii with smaller spatial deformation which characterizes the emission from 
the more strongly expanding later stages, superimposed by a non-trivial 
$K_\perp$ dependence driven by longitudinal boost dynamics.
The complexity of these competing processes demands a comprehensive study of 
two-photon interferometry and the corresponding HBT radii.

Bass {\it et al.} \cite{Bass:2004de} explored photon interferometry for 
central Au+Au collisions using ideal hydrodynamics coupled to a parton cascade 
to simulate the early preequilibrium emission \cite{Bass:2002fh}. 
They did not, however, investigate the azimuthal behavior of the photon HBT 
radii in non-central collisions. In this work we pioneer the exploration
of anisotropies of the photon HBT radii from non-central heavy-ion 
collisions, using the same (2+1)-dimensional hydrodynamic source 
exploited in our earlier work on photon elliptic flow 
\cite{Chatterjee:2005de,Heinz:2006qy,Turbide:2007mi,Gale:2008wf} to 
provide us with the thermal photon emission function. We examine photon 
HBT radii from both central and non-central collisions and compare them
with the corresponding pion HBT radii.

Before proceeding we must address several important differences between 
pion photon interferometry. First, for pions we were able to ignore spin. 
Photons carry spin 1 and possess two possible helicity states, and only 
photons with aligned helicities contribute to Bose-Einstein correlations 
through wave function symmetrization. Consequently, the strength of the photon 
correlator is reduced by a factor of 2 \cite{Slotta:1996cf}. The correlator 
between two photons with momenta $\bm p_a=\bm K{+}\frac{1}{2}\bm q$ and 
$\bm p_b=\bm K{-}\frac{1}{2}\bm q$, averaged over helicities, thus reads 
\begin{equation}
  C(\bm q,\bm K)\approx 1+\frac12\left|
  \frac{\int d^4x\,S(x,\bm K)e^{i\bm q\cdot\bm x}}
       {\int d^4x\,S(x,\bm K)}\right|^2\, ,
\label{eq:photonCorrelator}
\end{equation}
where we again applied the smoothness approximation, approximating the 
source functions at $\bm K\pm\frac12\bm q$ in the denominator of 
Eq.~\eqref{eq:GeneralCorrelator} by source functions at $\bm K$. A second, 
less obvious difference between 2-pion and 2-photon correlations arises from 
the fact that photons are massless which is addressed next.

\subsection{Failure of RMS variances for soft photons}
\label{sec4a}

It turns out that for soft photon pairs ($|\bm{K}|\lesssim 
1/R_\mathrm{source}$) both the on-shell and smoothness approximations 
discussed in Sec.~\ref{sec2a} become problematic. In this subsection, 
we consider these approximations in turn.

\subsubsection{On-shell approximation}
\label{sec4a1}
Consider the source RMS radii $R_o$ and $R_s$ defined in 
\eqref{eq:derivativeRMS},
\begin{eqnarray}
R^2_o = \langle \tilde{x}_o^2 \rangle 
      - 2\beta\langle\tilde{x}_o\tilde{t} \rangle 
      + \beta^2\langle \tilde{t}^2 \rangle, \quad
R^2_s = \langle \tilde{x}_s^2 \rangle, 
\label{eq:RMSphoton}
\end{eqnarray}
where for midrapidity pairs $\beta = \beta_\perp = K_\perp/K^0$. With 
the on-shell approximation $K^0=(E_a+E_b)/2\approx E_K$ (see 
Sec.~\ref{sec2a}) this becomes the pair velocity which for massless 
photons is $\beta=1$, independent of transverse pair momentum. The on-shell
approximation is valid for $q\ll 2K_\perp$, so it holds at $\bm{q}=0$
such that the RMS radii \eqref{eq:derivativeRMS} and \eqref{eq:RMSphoton}
continue to correctly describe the curvature of the correlation function
$C(\bm{q},\bm{K})$ at the origin. Equation \eqref{eq:RMSphoton} shows that 
for sources with non-zero emission duration $\langle\tilde t^2\rangle$ 
the outward and sideward RMS radii (and thus the curvature radii at the 
origin of the correlator in the corresponding $\bm{q}$ directions) differ
from each other {\em at all values of $K_\perp$}. This is contrary to the 
case of massive hadrons where the temporal contributions to the RMS radii
are suppressed at $K_\perp\to 0$ by powers of the pair velocity 
$\beta\to 0$ such that generically $R_s(K_\perp{=}0)=R_o(K_\perp{=}0)$
(for a discussion of possible exceptions see \cite{Tomasik:1998qt}).

The validity of the on-shell approximation is restricted to the region
$q\ll 2E_K$ which for midrapidity photon pairs shrinks to zero as
$K_\perp\to 0$. This implies that, for $K_l=0$ photons with small transverse 
pair momenta $K_\perp \lesssim 1/R_\mathrm{source}$ (where $R_\mathrm{source}$ 
characterizes the size of the emission region) the RMS radii 
\eqref{eq:derivativeRMS} describing the curvature of the correlator 
at $\bm{q}=0$ can no longer be expected to faithfully represent the 
width of the correlator. In other words, for photons the standard 
connection that, for Gaussian sources, relates the inverse width of 
the 2-photon correlator to the size of the photon-emitting region is 
broken for small pair momenta. Consequently, for soft photons the HBT 
radii extracted from a Gaussian fit to the 2-photon correlation function 
cannot be directly computed from the RMS widths of the photon emission 
function, {\em even if the latter is perfectly Gaussian}. At large 
$K_\perp\gg 1/R_\mathrm{source}$ this remains possible for Gaussian 
photon emission functions. However, since hydrodynamic photon emission 
functions are not sufficiently Gaussian (see below), the RMS shortcut 
should be avoided altogether and replaced by a Gaussian fit to the 
numerically computed 2-photon correlation function.

\begin{figure}
\begin{center}
\includegraphics[width=\linewidth]{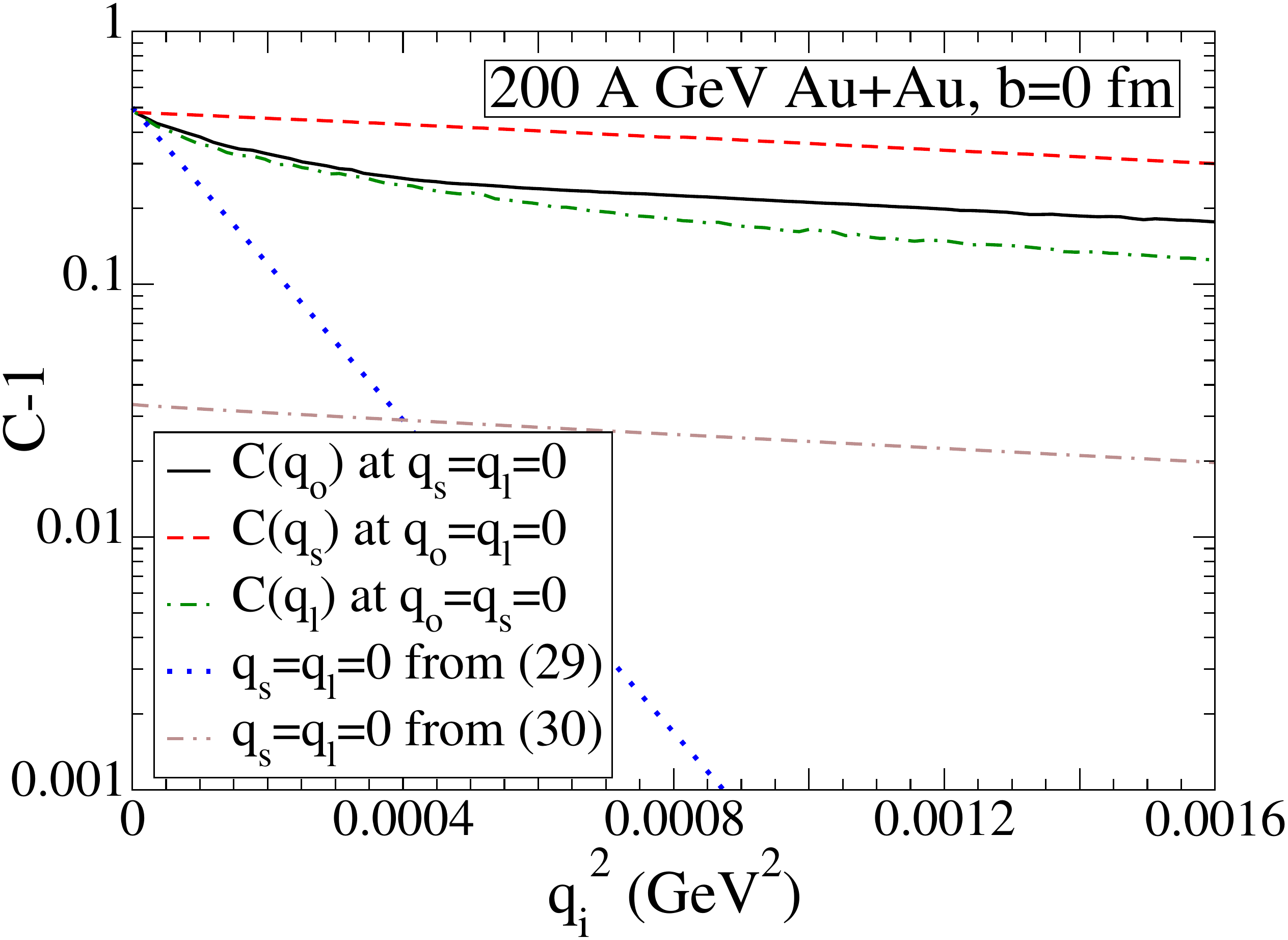}
\end{center}
\caption[1D slices of 2-photon correlation function, kT=0.01 GeV]{(Color 
online)
One-dimensional slices of the 2-photon correlation function $C(\bm{q},\bm{K})$
for soft photons from central 200\,$A$\,GeV Au+Au collisions, for midrapidity
pairs with $K_\perp=0.01$\,GeV/$c$. $C-1$ is plotted logarithmically
against $q_i^2$ ($i=o,s,l$) to facilitate visual extraction of the 
Gaussian ``width parameters'' from the slopes of the curves. The blue dotted 
and brown double-dash-dotted lines are computed from the RMS variances
of the source via Eqs.~\eqref{eq:firstTaylorCorrel} and 
Eq.~\eqref{eq:secondTaylorCorrel}, respectively. See text for discussion.
\label{fg:curvaturecompare}
}
\end{figure}

We illustrate this issue in Fig.~\ref{fg:curvaturecompare} by studying
the small-$q$ behavior of the 2-photon correlation function for soft 
midrapidity photon pairs with $K_\perp=0.01$\,GeV/$c$, computed numerically
from the hydrodynamic photon emission function. Shown are 1-dimensional 
slices of $C\,{-}\,1$ along the three axes $q_s$, $q_o$, and $q_l$ (from top 
to bottom), setting the two other $q$ components to zero. By plotting 
$C(\bm{q},\bm{K})-1$ logarithmically against $q_i^2$ one easily
identifies non-Gaussian features as deviations from linear behavior.
The ``Gaussian width parameters'' $R_i^2$ are given by the inverse slopes
of the lines. Clearly $R_l^2 > R_o^2 > R_s^2$ in the case shown. One sees 
that $C-1$ is a perfect Gaussian in the sideward, but not in the outward and 
longitudinal directions. Deviations from Gaussian behavior along the $q_l$
direction should be expected, since they are also seen in the 2-pion
correlation functions \cite{Frodermann:2006sp} and can be traced 
\cite{Chapman:1994ax,Wiedemann:1996ig} to the boost-invariant expansion 
dynamics of our hydrodynamic source. They extend over the entire $q_l$
range, i.e. $\ln[C(q_l^2)-1]$ can not be approximated by a straight line
anywhere (except for $q_l=0$). This is different for the outward correlator
$\ln[C(q_o^2)-1]$ (solid line in Fig.~\ref{fg:curvaturecompare}) which can 
be quite well described by two straight lines with different slopes at low 
and high $q_o^2$, with a smooth transition near $q_o^2=0.0004$\,(GeV/$c$)$^2
=(2K_\perp)^2$. Apparently, this change of slope is associated with the 
transition between the two limits discussed at the end of Sec.~\ref{sec2a}.
Such a change of slope is not seen for the 2-pion correlator which for
our hydrodynamical source is quite well described by a single Gaussian 
in the outward direction \cite{Frodermann:2006sp}.  

We can try to understand the behavior of the curves in 
Fig.~\ref{fg:curvaturecompare} analytically by writing (for $K_l{\,=\,}0$ 
pairs) $\beta=K_\perp/K^0(q)\equiv\beta(q)$, with the limits 
$\lim_{q\ll 2K_\perp}\beta(q)=1$ and $\lim_{q\gg 2K_\perp}\beta(q) =
2K_\perp/q$ for massless photons, and using this to express the exponent 
in Eq.~\eqref{eq:Gaussfit} in terms of the RMS variances 
\eqref{eq:derivativeRMS}, assuming a Gaussian source. In the limit of 
small relative momenta $q\ll 2K_\perp$ we obtain
\begin{widetext}
\begin{eqnarray}
\sum_{ij}q_iq_j\langle(\tilde x_i{-}\beta_i(q)\tilde t)
                      (\tilde x_j{-}\beta_j(q)\tilde t)\rangle =
\left\{
\begin{array}{cc}
q_s^2\langle\tilde x_s^2\rangle,\ & q_o=q_l=0,\\
q_o^2\left(\langle\tilde x_o^2\rangle - 2\langle \tilde x_o\tilde t\rangle 
         + \langle\tilde t^2\rangle\right),\ & q_s=q_l=0,\\
q_l^2\langle\tilde x_l^2\rangle,\ & q_o=q_s=0, 
\end{array}\right.
\label{eq:firstTaylorCorrel}
\end{eqnarray}
which is identical to the RMS variance calculations, setting 
$\beta =1$ for photons. For large relative momenta $q\gg 2K_\perp$
we find instead
\begin{eqnarray}
\sum_{ij}q_iq_j\langle(\tilde x_i{-}\beta_i(q)\tilde t)
                      (\tilde x_j{-}\beta_j(q)\tilde t)\rangle =
\left\{
\begin{array}{cc}
  q_s^2\langle\tilde x_s^2\rangle,\ & q_o=q_l=0,\\
  q_o^2\langle\tilde x_o^2\rangle 
- 4 K_\perp |q_o|\langle\tilde x_o\tilde t\rangle   
+ 4 K_\perp^2\langle\tilde t^2\rangle,\ & q_s=q_l=0,\\
  q_l^2\langle\tilde x_l^2\rangle,\ & q_o=q_s=0.
\end{array}\right.
\label{eq:secondTaylorCorrel}
\end{eqnarray}
\end{widetext}
We see that in this second limit factors of $\tilde t$ come without 
factors of $q_o$, but with factors of $K_\perp$ instead, and this 
changes the $q$-dependence of the correlator in $q_o$ direction. We note
that, if we had considered photon pairs with non-zero pair rapidity, an
analogous difference would have also appeared between the last lines in 
Eqs.~\eqref{eq:firstTaylorCorrel} for $q_o\ll 2K$ and 
\eqref{eq:secondTaylorCorrel} for $q\gg 2K$, where 
$K=\sqrt{K_\perp^2{+}K_l^2}$.

Figure~\ref{fg:curvaturecompare} assumes $K_l=0$, hence for a Gaussian
source $q_o^2 R_l^2=q_o^2 \langle \tilde x_l^2\rangle$ holds at all
values of $q$; the change of slope of the longitudinal correlator seen
in Fig.~\ref{fg:curvaturecompare} thus reflects deviations of the emission
function from a Gaussian form along the longitudinal direction. We will 
discuss these in more detail further below. The sideward HBT radius is 
always free from temporal factors, since $\beta_s\equiv0$ by definition.  
The sideward correlator in Fig.~\ref{fg:curvaturecompare} exhibits a 
perfectly constant slope, indicating that the hydrodynamic photon emission
function is well described by a Gaussian in $x_s$ direction.   

At large relative momentum $q_o$ the first term in the middle line of 
Eq.~\eqref{eq:secondTaylorCorrel} dominates, and the slope of $\ln(C-1)$ 
converges to $\langle \tilde x_o^2\rangle$, with no dependence on the 
emission duration left. This is interesting because it allows to
{\em separate the geometric from the temporal structure of the source}
with a single accurate measurement of the correlator at a fixed value
of $K_\perp$, by studying its different slopes at small and large $q_o$.
(We are aware, of course, of the experimental challenges of measuring 
thermal photon correlations at very small values of the pair momentum
$K_\perp$.) With massive hadron pairs this is impossible: For them 
Eq.~\eqref{eq:RMSphoton} is always a good approximation (as long as the 
source is sufficiently Gaussian), so a separation of geometrical
and temporal contributions requires measurements for different pair
velocities $\beta\sim K_\perp$, with the additional assumption that
the values of $\langle \tilde{x}^2 \rangle$, $\langle\tilde{x}\tilde{t}\rangle$
and $\langle \tilde{t}^2 \rangle$ don't themselves exhibit strong 
$K_\perp$-dependence.
 
According to Eq.~\eqref{eq:secondTaylorCorrel}, at large $q_o$ the outward 
slice of the correlator looks like a Gaussian with slope 
$\langle\tilde x_o^2\rangle$ and reduced intercept 
$\lambda=\frac{1}{2} e^{-4K_\perp^2\langle\tilde t^2\rangle}$.
At small relative momentum, its slope is characterized by 
$\langle\tilde x_o^2\rangle - 2\langle \tilde x_o\tilde t\rangle 
+ \langle\tilde t^2\rangle > \langle\tilde x_o^2\rangle$; this is 
bigger than the large momentum slope since for the hydrodynamic source 
both $-2\langle \tilde x_o\tilde t\rangle$ and $\langle \tilde t^2\rangle$ 
are positive. Eqs.~\eqref{eq:firstTaylorCorrel} and 
\eqref{eq:secondTaylorCorrel} thus explain qualitatively the change of 
slope along the outward direction seen in Fig.~\ref{fg:curvaturecompare}.

The dotted and double-dash-dotted lines in Figure~\ref{fg:curvaturecompare}
explore to what extent they also do so quantitatively. We see that they 
don't: the dotted line computed from Eq.~\eqref{eq:firstTaylorCorrel} for 
$q_o\ll 2K$ is much too steep whereas the double-dash-dotted line 
computed from \eqref{eq:secondTaylorCorrel} for $q\gg 2K$, while having
the correct slope, is much too low. The reason for this failure is subtle:
In the limit $K_\perp=0$, our photon emission function is independent of 
space-time rapidity $\eta$ since we use a hydrodynamic model with
longitudinal boost invariance. In this limit the variances 
$\langle \tilde z^2\rangle$ and $\langle \tilde t^2\rangle$ are infinite.
For small non-zero $K_\perp$ they are finite but very large. When this
happens, the expression \eqref{eq:Gaussfit}, together with 
Eq.~\eqref{eq:derivativeRMS}, can no longer be used because it relies
on a Taylor expansion of the correlator around $\bm{q}=0$, keeping only
the first non-trivial term and re-exponentiating the result. This 
works only for Gaussian source, but for $K_\perp\to0$ our source 
becomes very non-Gaussian along the $\eta$ direction, leading to a 
failure of this procedure in the terms containing $\langle \tilde z^2\rangle$ 
and $\langle \tilde t^2\rangle$. 

The Gaussian approximation continues to work for the first term 
$\sim\langle \tilde x_o^2\rangle$ in the middle 
Eq.~\eqref{eq:secondTaylorCorrel} which is why the double-dash-dotted 
line in Fig.~\ref{fg:curvaturecompare} correctly reproduces the slope of
the outward correlator slice (black solid line). In fact, this slope
is the same as for the sideward correlator slice (red dashed line),
indicating $\langle \tilde x_o^2\rangle=\langle \tilde x_s^2\rangle$
as should be the case for the central collisions studied in 
Fig.~\ref{fg:curvaturecompare}, due to azimuthal symmetry around the 
beam axis.

\begin{figure}
\begin{center}
\includegraphics[width=\linewidth]{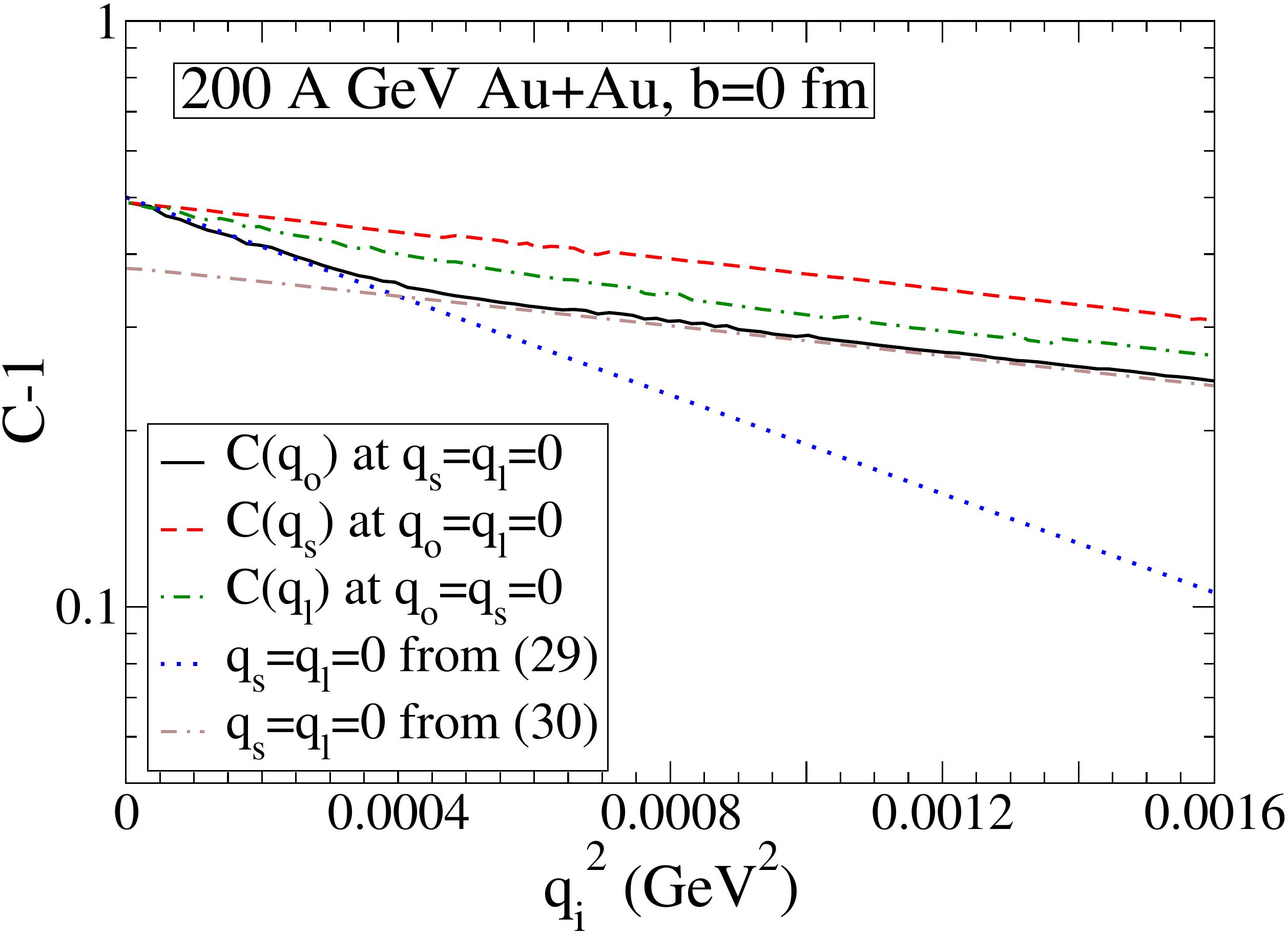}
\end{center}
\caption[1D slices of 2-photon correlation function, kT=0.01 GeV with $\eta$ 
correction]{(Color online)
Same as Fig.~\ref{fg:curvaturecompare} but for a modified emission function 
that has been multiplied by hand with a Gaussian cutoff factor $e^{-\eta^2}$
(see text for discussion).
\label{fg:curvaturecompare2}
}
\end{figure}

As a check for the above analysis we performed a calculation of the
correlator with a modified source where we multiplied the photon emission 
function from {\tt AZHYDRO} by hand with a Gaussian cutoff in $\eta$
direction, $S(x,K)\mapsto S(x,K)\times e^{-\eta^2}$. This simulates
a fireball with finite rapidity width $\Delta\eta=1/\sqrt{2}$. Now
the longitudinal and temporal variances $\langle \tilde z^2\rangle$ 
and $\langle \tilde t^2\rangle$ remain finite even for $K_\perp\to0$,
and the non-Gaussian features of the underlying boost invariant 
hydrodynamic emission function (which set in at larger $\eta$ values)
are suppressed. Fig.~\ref{fg:curvaturecompare2} shows that in this case
the slopes and magnitudes of the outward correlator slice are very well
reproduced by Eqs.~\eqref{eq:firstTaylorCorrel} and 
\eqref{eq:secondTaylorCorrel}. The problem in Fig.~\ref{fg:curvaturecompare}
is thus entirely due to the boost invariance of our hydrodynamic model.

As a corollary we note that, in contrast to hadrons whose rest 
mass always restricts their longitudinal homogeneity regions to
small subvolumes whose longitudinal extent is controlled by the inverse 
of the longitudinal expansion rate \cite{Heinz:1999rw,Wiedemann:1999qn},
low-momentum photons explore the full longitudinal size of the expanding
fireball. They can thus tell the difference between a longitudinally
finite and an infinite boost invariant source. Quantitatively reliable 
predictions of 2-photon correlations for pairs with small transverse 
momentum $K_\perp$ must therefore be based on realistic 3-dimensional 
evolution models that do not assume a boost invariant density 
distribution.

\subsubsection{Smoothness approximation}

In addition to the small $K_\perp$ breakdown of the on-shell approximation, 
the smoothness approximation also creates deviations which must 
be examined. The smoothness approximation is accurate as long as the 
curvature logarithm of the single-particle spectra is small 
\cite{Chapman:1994ax}. However, with the emission rates used here the 
photon spectra grow very rapidly at low momentum (see Fig~1 of 
\cite{Heinz:2006qy}), suggesting that in this region the smoothness 
approximation may also break down.

To explore the validity of the smoothness approximation we write the 
correlator \eqref{eq:GeneralCorrelator} as
\begin{equation}
\label{cor}
  C(\bm{q},\bm{K}) = 1\pm \frac{1}{g_s} D(\bm{q},\bm{K})
  \left|\frac{\int d^4x\,S(x,K)\,e^{i\,q\cdot x}}
             {\int d^4x\,S(x,K)}\right|^2
\end{equation}
where the last factor invokes the smoothness approximation 
\eqref{eq:TheoryDefinition} while deviations from this approximation
are captured by the ``correction factor''
\begin{equation}
\label{eq:smoothratio}
 D(\bm{q},\bm{K}) = \frac{\left|\int{d^4x\,S(x,\bm{K})}\right|^2}
 {\int{d^4x\,S(x,\bm{K}{+}\frac{\bm{q}}{2})}
  \int{d^4x\,S(x,\bm{K}{-}\frac{\bm{q}}{2})}}.
\end{equation}
When the smoothness approximation is valid, $D\approx1$.

We note that the correction factor $D$ is much harder to evaluate
than the ``smoothed'' correlator \eqref{eq:TheoryDefinition} which
can be efficiently computed by Monte Carlo integration with $S$ as
weight function. For this reason we usually evaluated the 3-dimensional
correlator $C(\bm{q},\bm{K})$ in the smoothness approximation and
used this as the basis for our 3-D Gaussian fits of the HBT radii.
Only for very small $K_\perp$ we also computed the correction factor
$D$ (see Fig.~\ref{fg:photoncent} below). In the present subsection 
we check the validity of the smoothness approximation by computing
the correction factor $D(\bm{q},\bm{K})$ only along the $q_o$,
$q_s$ and $q_l$ axes.
 
\begin{figure}
\includegraphics[width=\linewidth]{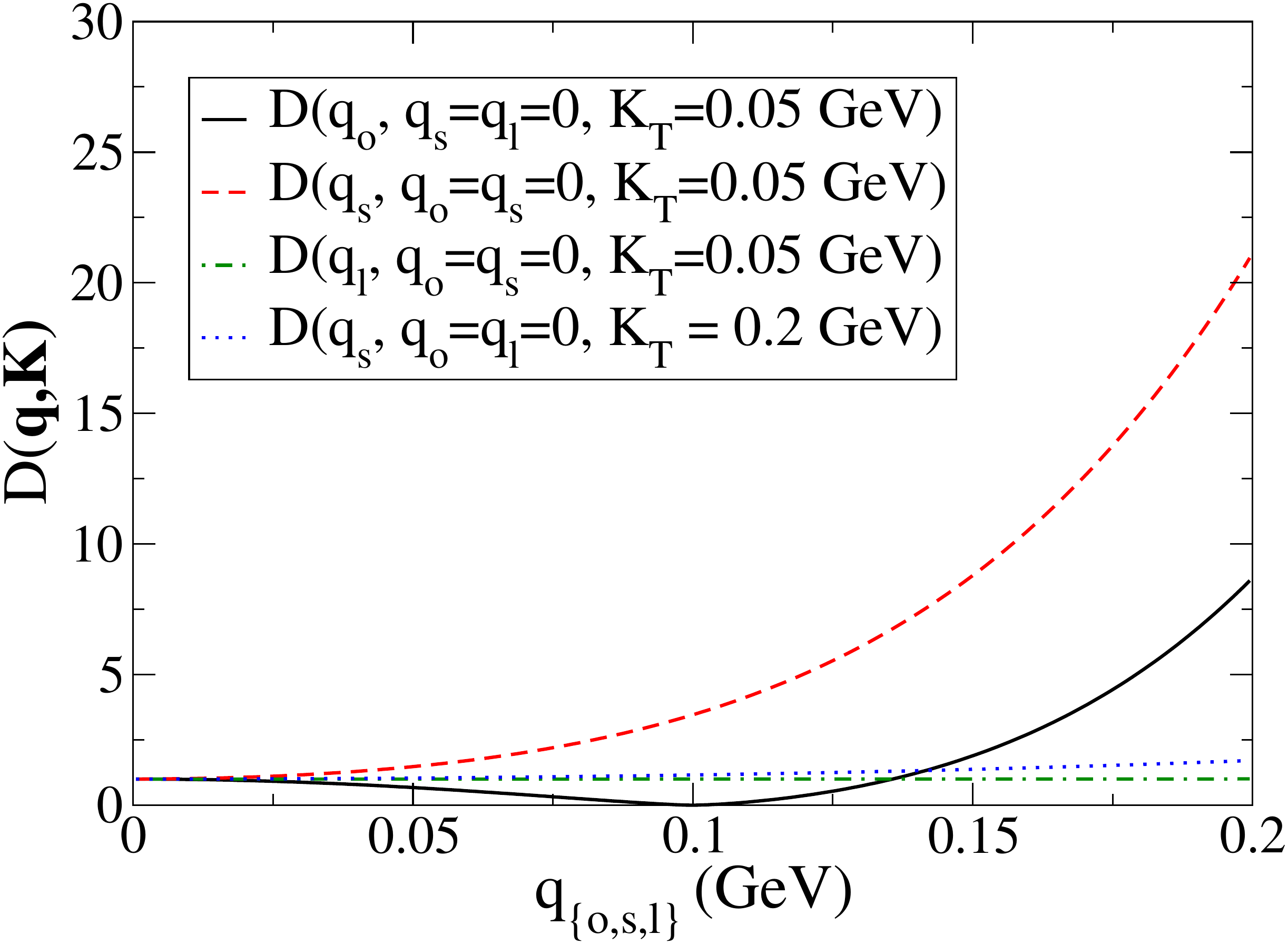}
\caption[Smoothness correction factor along the three slices of the thermal 
photon correlator ]{(Color online) The smoothness correction factor 
\eqref{eq:smoothratio} along the outward ($q_o$, black solid), sideward 
($q_s$, red dashed), and longitudinal ($q_l$, green dot-dashed) relative 
momentum axes, for central Au+Au collisions at 200\,$A$\,GeV and for
pair momentum $K_\perp = 0.05$\,GeV. For comparison the correction
factor along the sideward direction $q_s$ is also shown for pairs with 
larger transverse momentum $K_\perp=0.2$\,GeV (blue dotted line).
\label{fg:smoothratio}
}
\end{figure}

Figure~\ref{fg:smoothratio} shows the ``smoothness correction factor''
$D(\bm{q},\bm{K})$ as a function of relative momentum for photon pairs
with $K_\perp = 0.05$\,GeV. As the relative momentum increases, the ratio 
$D$ begins to deviate from unity for both the outward (solid black) and 
sideward (dashed red) slices. Along the $q_l$ direction (green dash-dotted) 
one sees almost no violation of the smoothness approximation. The 
characteristic momentum which sets the scale for violations of the 
smoothness approximation is again $\bm{q} = 2\bm{K}$, i.e. the same 
scale that characterizes the breakdown of the on-shell approximation.  
Due to the infinite longitudinal size of our boost invariant hydrodynamical
source, the single photon spectrum diverges at zero photon momentum;
as a result, the denominator of the ratio $D$ in Eq.~\eqref{eq:smoothratio} 
goes to infinity along the outward direction when $q_o = 2K_\perp$, 
generating a zero of $D(q_o,K_\perp)$ at this point. No such zero
occurs in the sideward and longitudinal directions where $\bm{K}\pm
\frac{\bm{q}}{2}$ never vanishes.

As we will see below (see Fig.~\ref{fg:photonslice}), the effects of the 
``smoothness correction factor'' on the 1-dimensional correlator slices
are largest in the $q_s$ direction. They matter only for small 
$K_\perp \lesssim 1/R_\mathrm{source}$. To illustrate this we show as 
blue dotted line in Fig.~\ref{fg:smoothratio} the correction factor
$D$ along the $q_s$ direction for pairs with $K_\perp=0.2$\,GeV (which
is larger than the inverse source radius in central Au+Au collisions).
In this case the deviations from unity are negligible over the entire
range where $C(\bm{q},\bm{K})$ deviates from unity.

\subsubsection{Non-Gaussian behavior along the longitudinal direction}

The non-Gaussian character of the longitudinal slice of the two-photon 
correlator in Fig.~\ref{fg:curvaturecompare} deserves some further 
discussion. It appears to be much stronger than previously observed
for pions: as we will see below, a 1D Gaussian fit to the longitudinal
correlation function leads to a $\sim20\%$ reduction of intercept parameter 
$\lambda_l$ whereas the analogous reduction for the two-pion correlator
is only about 5\% \cite{Frodermann:2006sp}. This suggests that the
non-Gaussian effects may be to some extent controlled by the rest mass 
of the emitted particles. 

To explore this issue, let us consider the following simple model
for an expanding thermalized Gaussian source~\cite{Wiedemann:1999qn}:
\begin{eqnarray}
\label{model}
  S(x,p) &=& K\,m_\perp\cosh(\eta)
  \exp\left({-\frac{p_\mu u^\mu}{T}}\right)
\nonumber\\ 
&\times&
  \exp\left({-\frac{\bm r^2}{2 \Delta \bm r}
                     -\frac{(\tau-\tau_0)^2}{2(\Delta\tau)^2}}\right).
\end{eqnarray}
The constant $K$ contains factors such as particle spin, fugacity, and 
the emission duration $\Delta \tau$. The emission function \eqref{model} 
has been studied extensively (see \cite{Wiedemann:1999qn} for a review). 
With our hydrodynamical model it shares the exact longitudinal boost 
invariance. By expanding the Boltzmann factor around $\eta=0$,
\begin{equation}
  \frac{p_\mu u^\mu}{T} \approx \frac{m_\perp}{T}
  \left(1+\frac{\eta^2}{2} + \ldots\right)\, ,
\end{equation}
and evaluating the emission function integrals for the RMS variances 
\eqref{eq:derivativeRMS} by saddle point integration one finds 
\cite{Wiedemann:1999qn,Makhlin:1987gm} 
\begin{equation}
  R_l^2 \approx \tau_0^2 \frac{T}{m_\perp},
\label{eq:GaussRl}
\end{equation}
where $\tau_0$ is the mean freeze-out time. This result suggests $T/m_\perp$ 
scaling for the longitudinal relative momentum ($q_l$) dependence of the 
two-particle correlation function. To isolate rest mass effects on the 
correlation function, we should study it as a function of the scaled
momentum $q_l\sqrt{T/m_\perp}$.

\begin{figure}
\includegraphics[width=\linewidth]{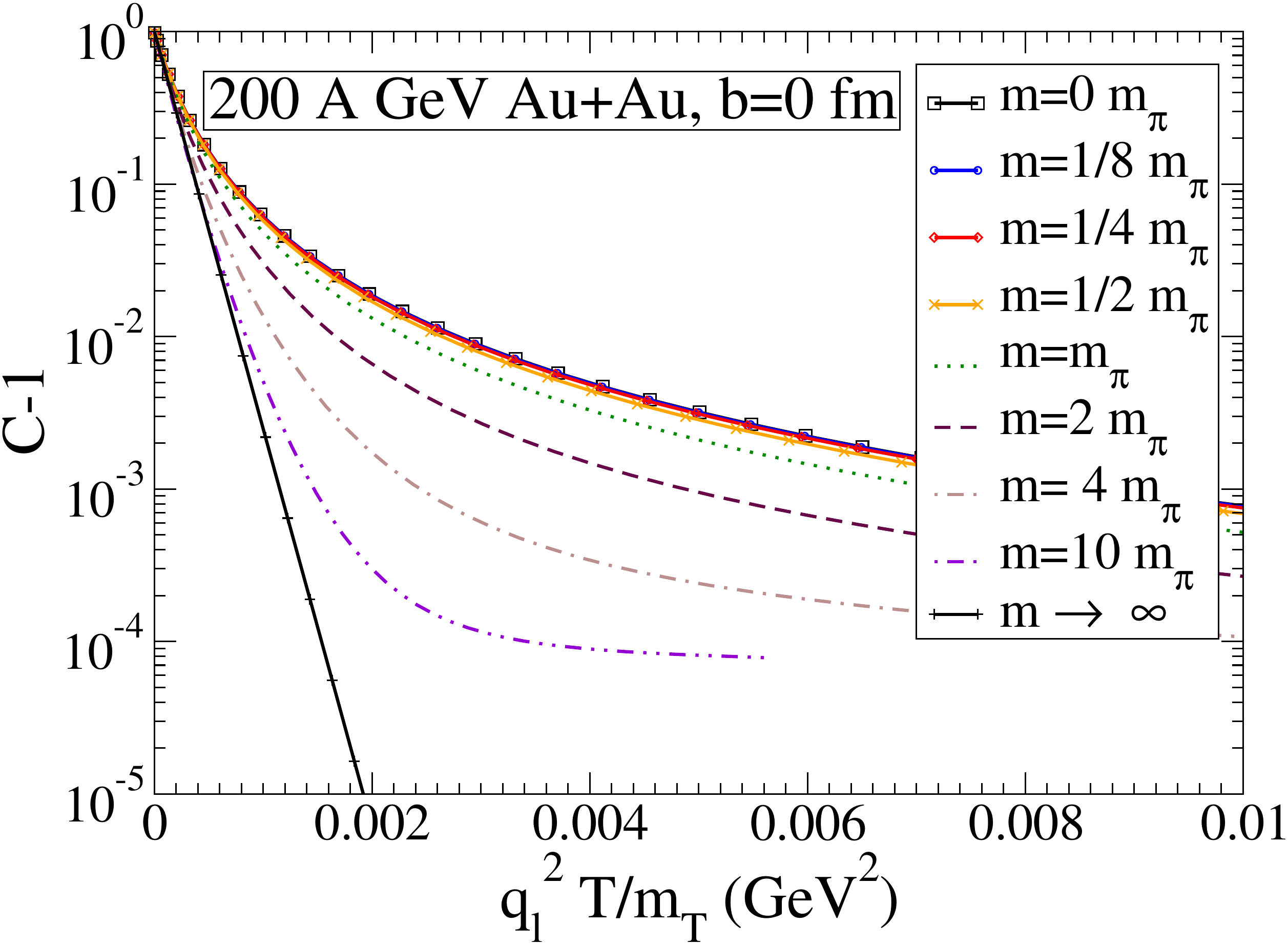}
\caption[Pseudo-pion mass effects on the pion correlation function]{(Color
online) Longitudinal slices of the two-particle correlation function
for spinless bosons (``pseudopions'') in the limit of zero pair momentum,
for different masses as given in the legend. The correlation functions 
were computed from the hydrodynamic pion emission function on the hadronic 
Cooper-Frye freeze-out surface, for 200\,$A$\,GeV Au+Au collisions at 
zero impact parameter, by varying the pion rest mass. They are plotted 
logarithmically against the scaled variable $q_l^2 \frac{T}{m_\perp}$. See 
text for discussion.  
\label{fg:pseudopion}
}
\end{figure}

Figure~\ref{fg:pseudopion} shows longitudinal slices of the two-particle
correlation function for spinless bosons of varying mass (``pseudopions''),
computed from the hydrodynamic emission function on the hadron freeze-out
surface for central Au+Au collisions at RHIC energies. The correlation
functions are plotted logarithmically against the scaled variable $q_l^2 
\frac{T}{m_\perp}$; the non-linear decrease of the correlators in this 
representation exposes their non-Gaussian character. At large $q_l$ the 
correlation functions are seen to drop much more slowly than expected from 
their (common) asymptotic slope at $q_l=0$. These deviations from a Gaussian
shape are strongest for low-mass particles and disappear for heavy 
particles. At infinite hadron mass, the correlator approaches a Gaussian 
function with a slope that agrees with the rest mass independent limiting
slope at $q_l=0$.

Figure~\ref{fg:pseudopion} explains why we see stronger non-Gaussian
effects in the longitudinal correlator for massless photons than for pions. 
Spot checks have shown that the non-Gaussian aspects are further 
enhanced in volume emission (as is the case for photons) relative to
Cooper-Frye surface emission (as simulated in Fig.~\ref{fg:pseudopion}
for ``pseudopions''). Since computing the volume emission function for 
photons is numerically more demanding than the surface emission function 
for pions, we do not present a systematic study of rest mass effects on 
volume emission.  

\subsection{Photon HBT radii}
\label{photonHBT}

In the remainder of this work we will concentrate on the transverse HBT
radius parameters for thermal photons. But since the 3-dimensional 
correlation function is non-Gaussian along the longitudinal direction,
this will in general contaminate the transverse HBT radii extracted
from a 3-d Gaussian fit even if the correlator is Gaussian in the
transverse directions \cite{Frodermann:2006sp}. We deal with this 
problem as follows:
  
\begin{figure*}
\begin{center}
\includegraphics[width=\linewidth]{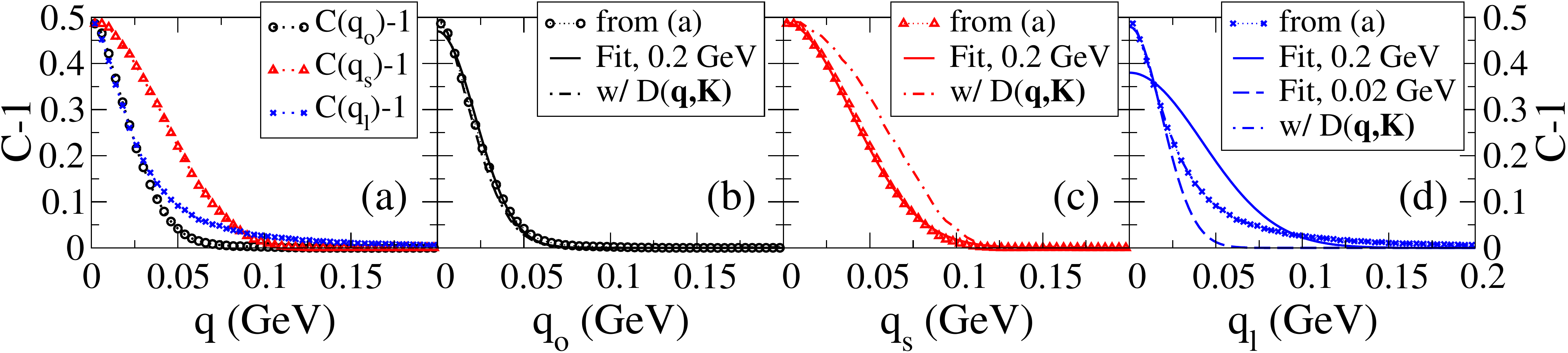}
\end{center}
\caption[1D slices of 2-photon correlation function, kT=0.05 GeV]{(Color
online) Slices of the photon correlation function $C(\bm{q})-1$ for 
central Au+Au collisions at $\sqrt{s_{NN}}$ = 200 GeV, for photon pairs 
with $K_\perp=0.05$\,GeV. Figure (a) shows slices of
the numerically calculated 3-dimensional hydrodynamic correlation 
function along the $q_o$ (circles), $q_s$ (squares) and $q_l$ (triangles)
directions. Note the strongly non-Gaussian shape of the $q_l$ slice.
The other three panels (b-d) show the individual slices superimposed with
1-dimensional Gaussian fits, using a fit range $q_i<q_\mathrm{max}$
with $q_\mathrm{max}=0.2$\,GeV/$c$ (solid lines) and (in the longitudinal
direction only) $q_\mathrm{max}=0.02$\,GeV/$c$ (dashed line). The dot-dashed 
line is the full correlation function taking into account the correction 
factor $D$ in Eq.~\eqref{eq:smoothratio} (see text for discussion) .  
\label{fg:photonslice}
}
\end{figure*}

Figure~\ref{fg:photonslice} shows 1-dimensional slices of the 3-dimensional
photon correlation function $C(\bm{q})-1$ along the $q_o,\,q_s,$ and $q_l$
directions, together with 1-dimensional Gaussian fits using a 1D analogue 
of the Gaussian fitting algorithm described in Sec.~\ref{sec2c}. The
fits (solid lines) were done in the range $0<q_i<0.2$\,GeV/$c$. One
notes the bad quality of this fit for the longitudinal slice 
(Fig.~\ref{fg:photonslice}d.), arising from the strongly 
non-Gaussian shape of the correlator
in this direction, and the associated significantly ($\sim 20\%$) reduced
intercept parameter. First principles tell us (and Fig.~\ref{fg:pseudopion}
confirms) that the correlator must be Gaussian near $q_l=0$. The dashed line
in Fig.~\ref{fg:photonslice} shows that by restricting the $q_l$ fit range 
to $q_l<0.02$\,GeV/$c$ (i.e. a 10 times smaller window) we can largely
cut out the non-Gaussian effects and obtain a fit with a reasonable 
intercept parameter close to 0.5 which is not in conflict with the
similar intercept parameters preferred by the 1D Gaussian fits in $q_s$ 
and $q_o$ directions. 

The dot-dashed lines show slices of the thermal photon correlation function 
Eq.~\eqref{eq:GeneralCorrelator} which is obtained by multiplying the 
numerically computed curves in Fig.~\ref{fg:photonslice}a, which use the 
smoothness approximation Eq.~\eqref{eq:TheoryDefinition}, by the correction 
factor Eq.~\eqref{eq:smoothratio}.  Although both the outward and 
sideward slices are highly sensitive to deviations from the ``smoothed'' 
correlator at low relative momentum, the characteristic momentum 
$\bm{q}=2\bm{K}$ is sufficiently in the tail of the correlator that only the 
sideward slice exhibits these deviations.  The deviations along $q_o$ only 
drive the correlator to zero faster and only in the tail when the correlator 
is sufficiently small.

In the following, we will therefore show transverse HBT radii that have 
been obtained from 3D Gaussian fits to the numerically calculated photon
correlation functions $C(\bm{q})$ using a restricted $q_l$ fit range
$|q_l|<0.02$\,GeV in the longitudinal direction but including in the 
transverse directions the full range of $q_o,\,q_s$ values where $C$ 
deviates significantly from unity. This procedure is necessary if one
wants to obtain meaningful values for the azimuthal oscillation amplitudes
of the transverse HBT radii in non-central collisions (see Sec.~\ref{osc}).
If these are extracted from 3D fits with unrestricted $q_l$ range,
the non-Gaussian features in $q_l$ lead to azimuthally oscillating 
intercept parameters $\lambda(\Phi)$ which contaminate and unphysically 
distort the azimuthal oscillations of the HBT radii in ways that do
not reflect the azimuthal dependence of the size and shape of 
the emission function.
 
From the 3D Gaussian fits with restricted $q_l$ range we also extract 
longitudinal HBT radii. Obviously, these reflect the curvature of the
longitudinal correlator near $q_l=0$ and do not necessarily provide an
accurate description of its longitudinal width. The latter is more
accurately given by 1D Gaussian fits along $q_l$, and we will therefore 
show these 1D $R_l$ values for comparison. 

\subsection{Central collisions}
\label{HBTCent}

\begin{figure}
\begin{center}
\vspace*{-5mm}
\includegraphics[width=\linewidth]{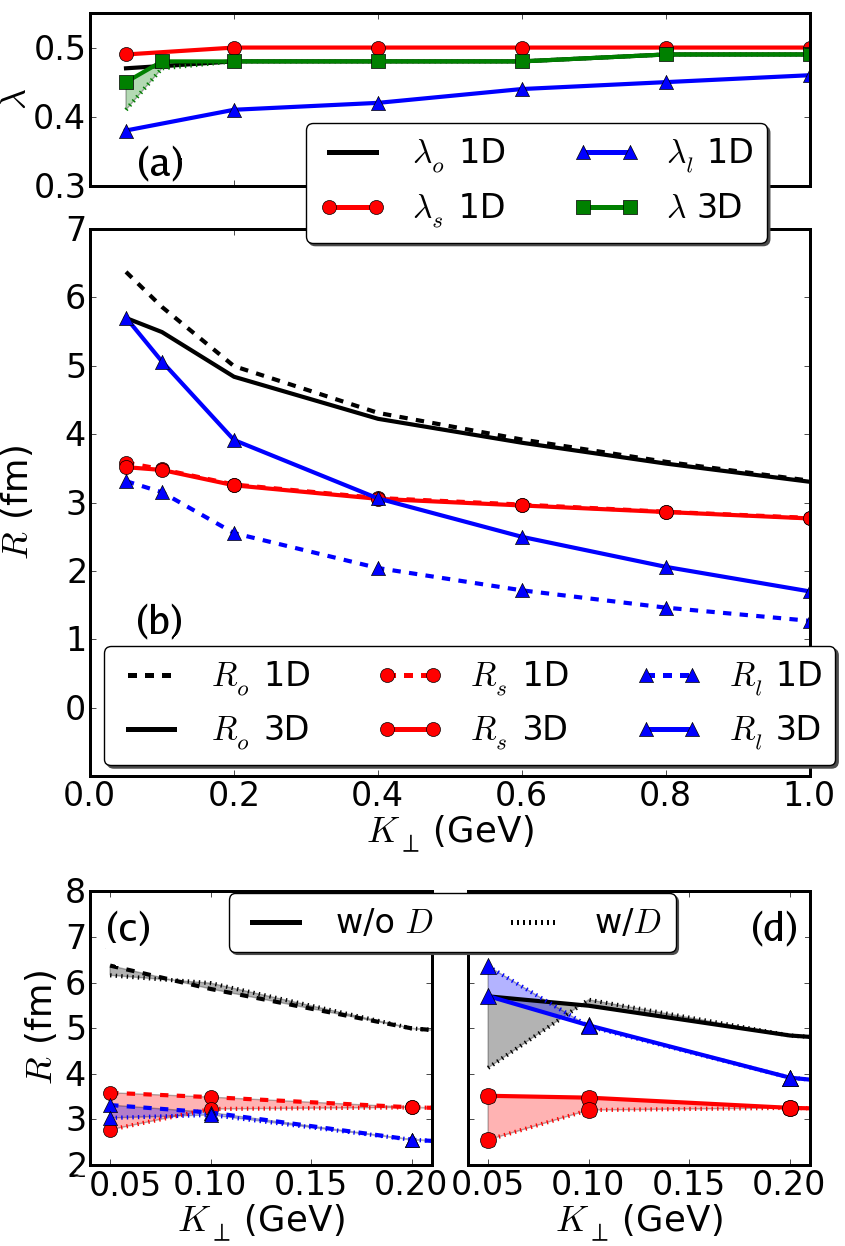}
\end{center}
\caption[Photon HBT radii for Au+Au @ 200 GeV, b=0 fm]{(Color online)
{$K_\perp$-dependence of correlation strength parameter $\lambda$ (a) and 
HBT radii (b-d) from 1- and 3-dimensional Gaussian fits of the hydrodynamic 
2-photon correlation function for central Au+Au collision at RHIC energies. 
For the 3D fits we used $q_{o,s}^{\rm max}=0.2$\,GeV and $q_{l}^{\rm max} 
= 0.02$\,GeV; the 1D longitudinal fits were done with $q_{l}^{\rm max} = 
0.2$\,GeV. (b): HBT radii from 1D and 3D Gaussian fits to the 
smoothness-approximated correlator \eqref{eq:TheoryDefinition}. (c):  
Low-$K_\perp$ blowup of the 1D fit radii from (b) (dashed lines), compared
with 1D fit radii for the smoothness-corrected correlator \eqref{cor} 
(dotted lines). (d): Low-$K_\perp$ blowup of the 3D fit radii from (b) 
(solid lines), compared with 3D fit radii for the smoothness-corrected 
correlator \eqref{cor} (dotted lines).}   
\label{fg:photoncent}
}
\end{figure}

Figure~\ref{fg:photoncent}b shows photon HBT radii from the hydrodynamic
model for central 200\,$A$\,GeV Au+Au collisions as a function of photon
pair momentum $K_T$, obtained from Gaussian fits to the numerically 
computed correlation function as described above. In out- and sideward 
directions the 1D and 3D HBT radii are almost identical, except at small
$K_\perp$, reflecting an approximately Gaussian shape of the correlator
in transverse directions. As discussed in Sec.~\ref{sec4a}, the outward
two-photon correlator develops non-Gaussian features at small $K_\perp$,
and the strong non-Gaussian features in the longitudinal direction (visible
in the large discrepancy between the 1D and 3D longitudinal fit radii $R_l$) 
get also worse at small $K_\perp$. Both effects drive the common intercept
parameter $\lambda$ for the 3D fit (squares in Fig.~\ref{fg:photoncent}(a))  
below 0.5, especially at low $K_\perp$, in spite of the restriction of the 
longitudinal fit range to $q_l<q_l^{\rm max}= 0.02$\,GeV. For the 1D 
longitudinal fit which uses a 10 times larger longitudinal fit range, 
$\lambda$ shows a nearly 25\% deviation from the expected value of 
0.5 as $K_\perp \to 0$. This is our motivation for restricting the 
longitudinal fit range in the 3D fits, in order to minimize contamination 
of the transverse radii extracted from the 3D fit that stems from a 
reduction of the common intercept parameter exclusively driven by 
strongly non-Gaussian features along the longitudinal direction. 

The dotted curves in Figures~\ref{fg:photoncent}(c) (1D radii) and 
\ref{fg:photoncent}(d) (3D radii) show the changes to the Gaussian fit 
radii when the ``smoothness correction factor'' \eqref{eq:TheoryDefinition} 
is applied to the correlator, as shown in Eq.~\eqref{cor}. For the 1D fit 
radii the smoothness correction factor has almost no effect along the
outward and longitudinal directions; in the sideward direction, the 
broadening of the correlator by the smoothness correction factor, shown 
as the dash-dotted line in the lower left panel of Fig.~\ref{fg:photonslice}, 
leads to a 25\% decrease of the 1D sideward radius $R_s$ at 
$K_\perp=0.05$\,GeV. 

The 3D fits, however, show large smoothness correction effects for {\em 
all three\,} extracted radii at small $K_\perp$ where the factor $D$ in 
Eq.~\eqref{eq:smoothratio} introduces non-Gaussian deformations also in 
the side- and outward directions (see Fig.~\ref{fg:smoothratio}). At 
$K_\perp=0.05$\,GeV these, together with the inherent non-Gaussian 
structure in $q_l$-direction already at the smoothness-approximated level, 
cause a reduction of the common 3D intercept parameter by almost 20\% below 
its ideal value of $\frac{1}{2}$; this, in turn, further distorts the 
three Gaussian fit radii. These distortions happen only at very small 
pair momentum; already at $K_\perp=0.1$\,GeV the smoothness-correction 
effects are almost completely gone. 

The most striking feature of Fig.~\ref{fg:photoncent}(b) is the strong
divergence between $R_s$ and $R_o$ as $K_\perp\to0$. This is due to
the temporal contributions (in particular the emission duration 
contribution) to $R_o^2$ which, for massless photons, are not suppressed
by pair velocity factors in the limit $K_\perp\to 0$ as is the case
for hadrons. As a result, for soft photon pairs the outward HBT radius
exceeds the sideward one by more than 50\%. We have checked that this 
effect disappears, i.e. $\lim_{K_\perp\to0}(R_o-R_s)=0$ for the 3D Gaussian 
radii, if we give the photons a small nonzero mass.

\begin{figure}
\begin{center}
\includegraphics[width=\linewidth]{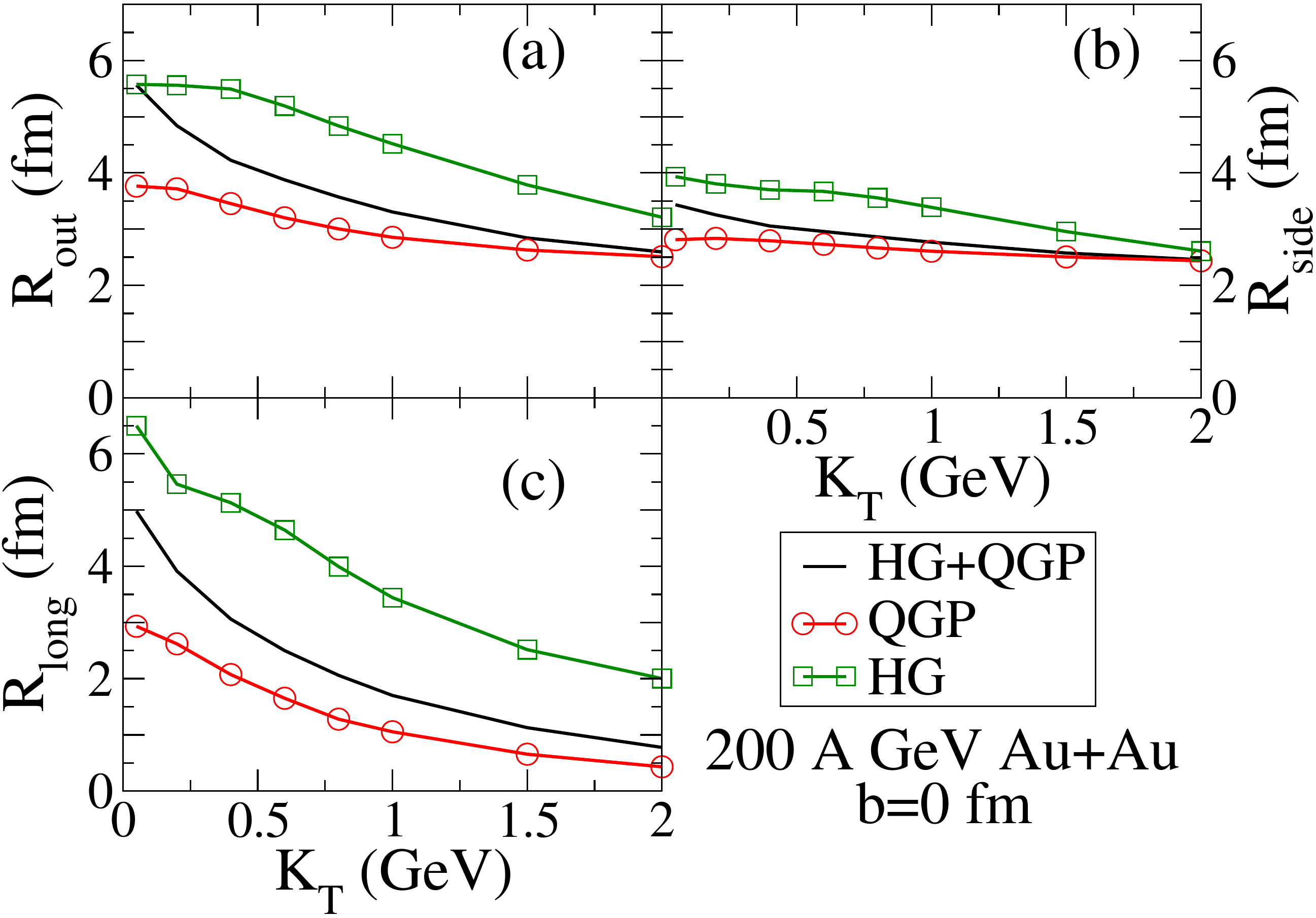}
\end{center}
\caption[Photons: HG and QGP extract HBT radii]{(Color online) 
3D Gaussian HBT radii as a function of transverse pair momentum for
the photon emission functions describing radiation from the quark-gluon 
plasma (QGP), from the hadron gas stage (HG), and the total emission.
\label{fg:HGQGPphotoncent}
}
\end{figure}

In Figure~\ref{fg:HGQGPphotoncent} we separate the photon emission function
into two contributions, corresponding to radiation from the quark-gluon 
plasma and hadron gas phases. Not surprisingly, the HG HBT radii indicate 
a larger source for the hadronic phase than for the QGP phase whose size
is reflected in the QGP HBT radii. The measured photon HBT radii (QGP+HG)
compromise between these values, reminding us that photons are 
emitted throughout the fireball evolution. At large $K_\perp$, the
transverse HBT radii from the total source approach those from the QGP
phase, reflecting the fact that at large $K_\perp$ QGP radiation dominates
the single photon yield. The transverse photon HBT radii from the HG phase 
are not too different in magnitude from the corresponding pion radii, 
except for the emission duration effects in $R_o$ at small $K_\perp$ for 
photons. The difference between $R_o$ and $R_s$ at small $K_\perp$ is 
smaller for the QGP than for the HG emission function, indicating shorter 
effective emission durations for QGP photons compared to hadronic photons.

The longitudinal photon HBT radius from the HG phase is significantly 
smaller than the pionic one, but this is at least partially due to the 
stronger non-Gaussian effects in the longitudinal correlation function 
for photons and our restriction of the $q_l$ fit range in the 3D 
fits. The QGP photon radii are significantly smaller than the HG radii, 
especially for the longitudinal radius \cite{Bass:2004de}, reflecting 
the strong longitudinal expansion velocity gradient at early time 
\cite{Wiedemann:1999qn,Makhlin:1987gm} when most of the QGP radiation 
is emitted (see also Eq.~\eqref{eq:GaussRl}), and acerbated by the strong 
non-Gaussian nature along the longitudinal direction. (We note that for photons
we find $R_l < R_o$ at all values of $K_\perp$ whereas the inverse 
relation $R_o < R_l$ holds for pions \cite{Frodermann:2006sp}.) 

\subsection{Non-central collisions}
\label{HBTNonCent}
\subsubsection{Photon emission functions}
\label{EF}
 
The results from the previous subsection for central collisions have shown 
that the photon HBT radii can only be properly understood if one keeps in 
mind that photons are emitted from all stages of the expanding fireball.
This is even more true for non-central collisions and for the azimuthal
oscillations of the HBT radii in this case. In this subsection we show
that at nonzero $K_\perp \sim 1$\,GeV, the photon emission functions
exhibit qualitatively different characteristics from those we are familiar
with from pions \cite{Heinz:2002sq}.   

\begin{figure}
\begin{center}
\includegraphics[width=\linewidth]{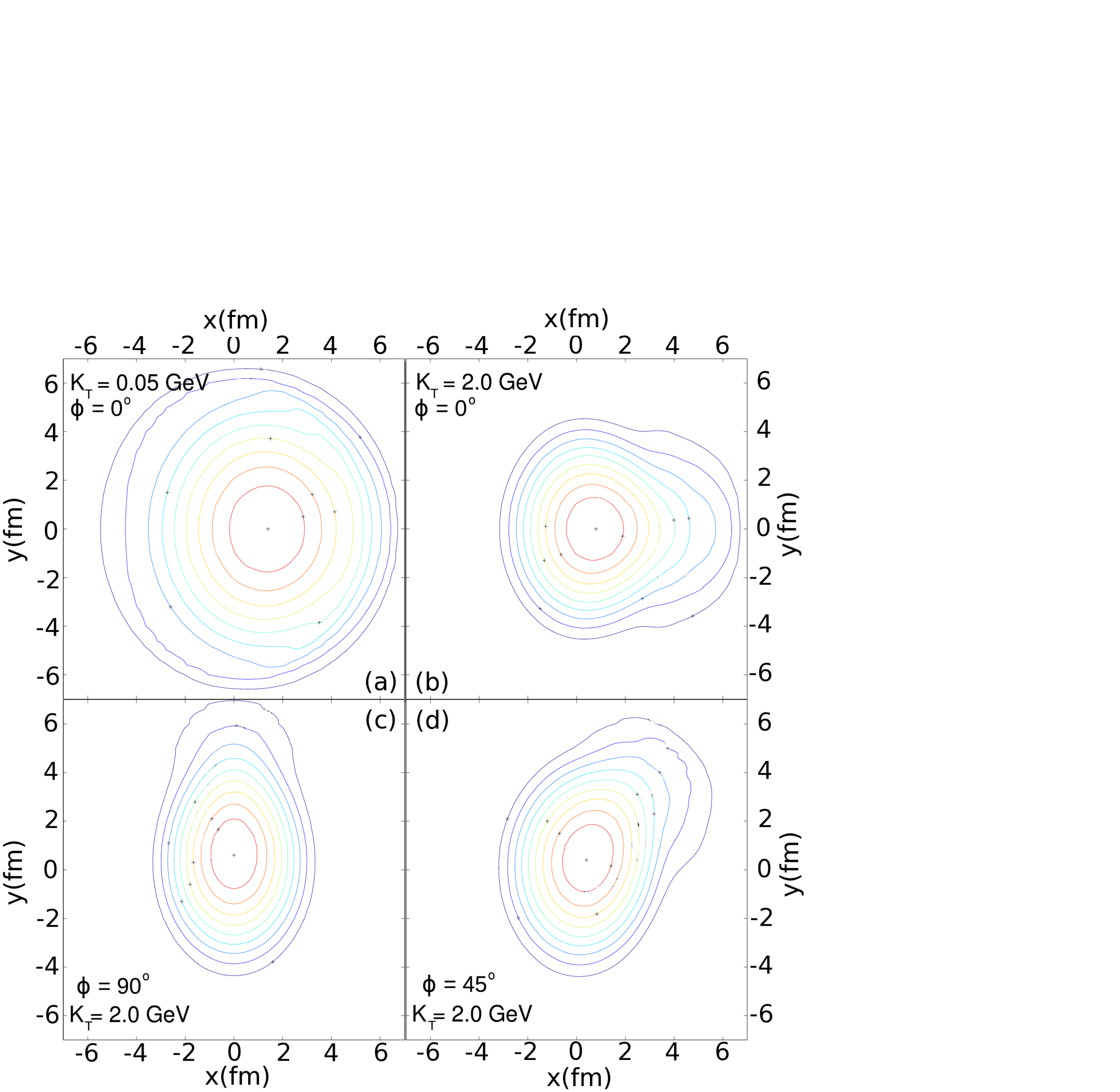}
\end{center}
\caption[Photon emission function at kT = 2.0 GeV at various phi]{(Color
online) Photon emission function $S(x,y)$ at $K_\perp=0.05\,\mathrm{GeV}
\approx 0$, $\Phi=0$ (upper left panel) and at $K_\perp = 2.0$\,GeV for
three emission directions, $\Phi = 0^\circ,\,45^\circ,$ and $90^\circ$
(other three panels), for 200\,$A$\,GeV Au+Au collisions at $b=7$\,fm.
The emission functions are normalized to 1 at their maxima; contours are
drawn in 10\% intervals from this maximum value.
\label{fg:photonSXY2}
}
\end{figure}

Figure~\ref{fg:photonSXY2} shows contour plots of the normalized
emission function $S(x,y;K_\perp)$, integrated over $z$ and $t$,
for midrapidity photons at $K_\perp=0.05$ and 2\,GeV/$c$. Since 
$K_\perp=0$ photons are emitted from essentially everywhere in 
the source at any time, but the space-time volume covered by the 
hadronic phase exceeds that of the QGP at RHIC energies
\cite{Chatterjee:2005de}, the contour lines in Figure~\ref{fg:photonSXY2}a 
indicate the overall size and shape of the momentum-integrated photon 
emission function, with a stronger weight on the late evolution stages. 
The slight distortion of the contours towards the right (positive $x$ 
direction) reflects the nonzero pair momentum $K_\perp=0.05$\,GeV/$c$ 
in $\Phi=0$ direction. Emission of such photons is dominated by cells 
which flow hydrodynamically in $+x$ direction with a flow velocity that 
has just the right magnitude to boost the peak of the Boltzmann 
distribution to this nonzero $K_\perp$ value. Due to the dominance
of the HG phase for the emission of low-$K_\perp$ photons
\cite{Chatterjee:2005de,Heinz:2006qy}, this reflects the strong radial 
flow in the {\em late hadronic stage} of the fireball.

Photons with larger $K_\perp=2$\,GeV/$c$, on the other hand, are
mostly emitted from the early QGP stage \cite{Chatterjee:2005de,Heinz:2006qy}.
At this point, radial flow is weak, so we should naively expect the photon
emission function to peak near the center of the fireball, with little
distortion by radial flow. This is in strong contrast to pions
which are only emitted at the end of the evolution from the
hadronic freeze-out surface and thus always feel the full radial
flow of the late hadronic fireball. For large $K_\perp$ values, hadronic 
emission functions are strongly peaked near the edge of the fireball,
forming narrow crescent-shaped slivers that straddle the surface of
the fireball \cite{Heinz:2002sq} since it is in these regions that 
one finds the strongest radial flow which most efficiently boosts the 
local thermal distribution with temperature $T_\mathrm{f}$ towards large 
$K_\perp$ values. 

The three $K_\perp=2$\,GeV/$c$ panels (Fig.~\ref{fg:photonSXY2}(b-d)) show 
something quite different: For large $K_\perp$, the photon emission 
function peaks near the center of the fireball (at early times, as 
studies of its $t$-dependence have revealed), but it exhibits a 
``bulge'' in the direction of the emitted photons that reflects a 
contribution from later times when the matter is in the hadronic phase 
and boosted by strong radial flow. Even though the hadronic phase
radiates at much lower temperature ($\langle T_\mathrm{HG}\rangle
\sim 150$\,MeV while $\langle T_\mathrm{QGP}\rangle$ is almost twice as high),
which strongly suppresses its photon emission rate, the larger
space-time volume covered by the HG phase partially compensates for
that loss in rate and makes hadronic photon emission sufficiently
competitive at $K_\perp=2$\,GeV/$c$ to give a visible flow-boosted
contribution to the photon emission function. The hadronic component 
is not strong enough to collectively ``squeeze'' the photon emission 
function towards the edge of the fireball (as was the case for pions),
but it distorts it visibly into the direction of the emitted photon 
pair. This parallels the study of photon elliptic flow which showed 
\cite{Chatterjee:2005de,Heinz:2006qy} that, even though at 
$K_\perp=2$\,GeV/$c$ the single-photon yield is dominated by QGP 
radiation, the stronger elliptic flow of the HG photons still 
controls the total photon $v_2$, and one must go to significantly 
larger transverse momenta before the elliptic flow of photons provides 
an uncontaminated view of the earliest QGP stage.

Were it not for the hadronic ``bulge'', the $K_\perp=2$\,GeV/$c$ 
emission functions shown in Fig.~\ref{fg:photonSXY2} would indeed 
provide a clean reflection of the original source eccentricity in 
non-central collisions. The emission function in the 
Fig.~\ref{fg:photonSXY2}b-d (dominated by QGP radiation) is clearly more 
eccentric than the one in the Fig.~\ref{fg:photonSXY2}a (which is dominated 
by hadronic radiation). What Fig.~\ref{fg:photonSXY2} teaches us is that the 
full story is more complex than suggested by this naive view, and that any 
interpretation of azimuthal oscillations of the photon HBT radii
must properly account for the two-component structure of the photon 
emission function which is generated by the interplay between 
non-boosted QGP and flow-boosted hadronic photon emission, with 
relative weights that depend on $K_\perp$.

\subsubsection{Azimuthal oscillations of photon HBT radii}
\label{osc}
 
\begin{figure}
\begin{center}
\includegraphics[width=\linewidth]{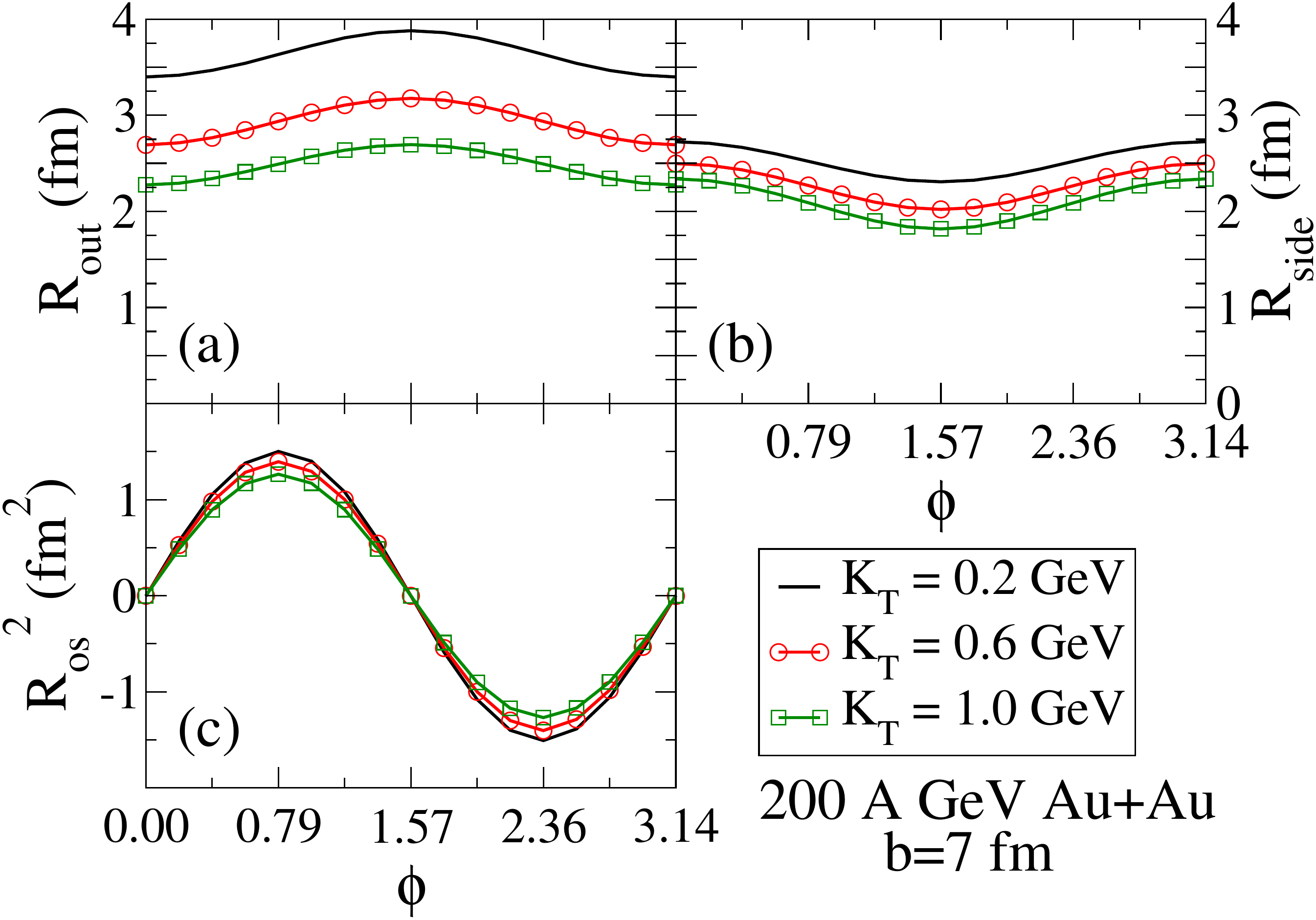}
\end{center}
\caption[Photon: b = 7 fm, radii vs. azimuthal angle]{(Color online)
Azimuthal oscillations of the transverse photon HBT radius parameters 
$R_o$ (upper left), $R_s$ (upper right), and $R_{os}^2$ (lower left),
for photon pairs with $K_\perp = 0.2,\, 0.6,$ and 1.0\,GeV/$c$ from
semi-peripheral 200\,$A$\,GeV Au+Au collisions at $b=7$\,fm.
\label{fg:photonosci1}
}
\end{figure}

Figure~\ref{fg:photonosci1} shows the azimuthal oscillations of the 
transverse photon HBT radius parameters in semi-peripheral Au+Au 
collisions at RHIC. The HBT radii were extracted from a 3D Gaussian fit
to the correlation function, with restricted longitudinal fit range
$q_o<0.02$\,GeV/$c$ as explained above. Qualitatively, the features
seen in Fig.~\ref{fg:photonosci1} agree with those for pions shown in 
Fig.~\ref{fg:pionHBTNC}. On a more quantitative level, one observes 
important differences: the sideward radius is smaller for photons,
reflecting preferred emission from the smaller, hotter fireball 
interior and earlier times, while the relative oscillation amplitude
is larger, reflecting the stronger source eccentricity at early times.  
The $K_\perp$-dependence of the azimuthally averaged $R_s$ value is 
weaker for photons than for pions; the likely reason is weaker radial 
flow at earlier times when a large fraction of the photons are emitted.
The $K_\perp$-dependence of $R_o$ is much stronger for photons than for
pions, mostly due to the much larger photon $R_o$ radii at small $K_\perp$
resulting from the emission duration contribution which at low $K_\perp$ 
is suppressed for pions but not for photons (see discussion in 
Sec.~\ref{sec4a}). 

Interestingly, the $R_{os}^2$ cross term for photons oscillates with an 
amplitude that is almost independent of $K_\perp$ and even bigger at 
small $K_\perp$ than at large $K_\perp$.  Fig.~\ref{fg:photonosci1}c exhibits 
almost pure $\sin(2\Phi)$ oscillations 
for the $R_{os}^2$. The relationship \cite{Lisa:2000ip,Heinz:2002sq}
\begin{eqnarray}
\label{Ros}
  R_{os}^2 &=& \cos(2\Phi)\langle\tilde{x}\tilde{y}\rangle
           + \sin(2\Phi)\frac{\langle \tilde{y}^2-\tilde{x}^2\rangle}{2}\\
           &\hspace{0.5cm}+&\beta(q) 
	   \Bigl(\langle\tilde{x}\tilde{t}\rangle\sin\Phi
           - \langle\tilde{y}\tilde{t}\rangle\cos\Phi\Bigr),\nonumber
\end{eqnarray}
which holds for emission functions whose dependence on $x,y,t$ is Gaussian 
and should thus be reasonably accurate for our source, indicates that the
larger $\sin(2\Phi)$ oscillation amplitude for photons at low $K_\perp$ 
is caused by a larger source eccentricity for low-momentum photons than
for low-momentum pions. This is due to the significant weight of early 
photon emission even at zero transverse pair momentum.

For pions it was found that $K_\perp=0$ pion pairs are emitted from the 
entire fireball, and that the normalized azimuthal oscillation amplitude 
of the sideward radius $R_s$ can thus be used to measure the spatial
eccentricity of the fireball at pion freeze-out \cite{Retiere:2003kf}.
In the same spirit we can try to extract the fireball eccentricity
from the normalized azimuthal oscillations of the photon sideward radius
at $K_\perp$. The sideward radius is the signature of choice since
it is a purely geometric observable, uncontaminated by temporal 
contributions (see Eqs.~\eqref{eq:firstTaylorCorrel} and 
\eqref{eq:secondTaylorCorrel}). The discussion above indicates, 
however, that photons even at $K_\perp=0$ are emitted more often from
the hot fireball center at early times than from the cooler periphery
at later times so the normalized azimuthal $R_s$ oscillation for photons
will measure the effective fireball eccentricity at earlier times than
for pions. This is, of course, exactly what we hoped to obtain. The 
only unexpected aspect is that this would work even for $K_\perp=0$
where we did not anticipate early emission to play quite as important
a role as we now see.

The contour plots in Fig.~\ref{fg:photonSXY2} show that the space-time 
character of photon emission differs from pions even more strongly at 
larger $K_\perp$ where photon emission is even more strongly concentrated
at early times close to the fireball center whereas pion emission is
almost completely surface dominated and concentrated to a thin sliver
near the fireball edge \cite{Heinz:2002sq}. This makes a geometric 
interpretation of the normalized $R_s$ oscillation amplitude in terms
of spatial eccentricity of the photon emission function possible even
at larger $K_\perp$ values. By going to larger $K_\perp$, we can thus
measure with photons the fireball eccentricity at even earlier times,
by analyzing their normalized $R_s$ oscillation amplitudes as a
function of $K_\perp$. (For pions the straightforward geometric 
interpretation of this observable is lost for $K_\perp\gg 0$ 
\cite{Retiere:2003kf}.)

\begin{figure}
\begin{center}
\includegraphics[width=\linewidth]{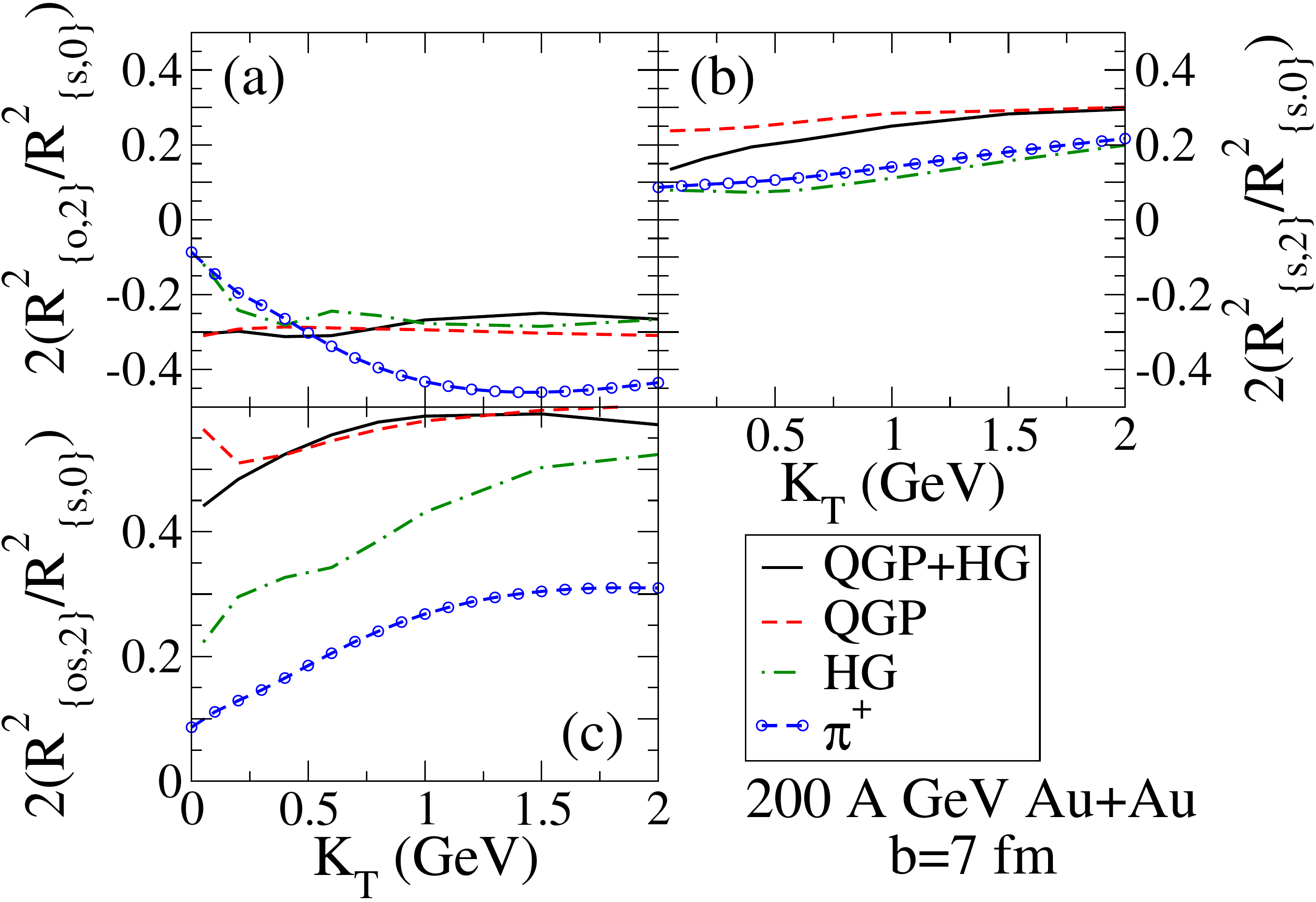}
\end{center}
\caption[Photon: b = 7 fm, oscillation amplitudes vs. kT]{(Color online)
Normalized oscillation amplitudes of the photon (left column) and pion 
(right column) HBT radii from 200\,$A$\,GeV Au+Au collisions at $b=7$\,fm,
as a function of transverse pair momentum.
\label{fg:photonAmps}
}
\end{figure}

In Fig.~\ref{fg:photonAmps} we show the normalized oscillation amplitudes
\cite{Heinz:2002au} which, following Reti\`ere and Lisa 
\cite{Retiere:2003kf}, can be expressed through the following normalized 
2$^{\rm{nd}}$ order azimuthal Fourier components of the HBT radius 
parameters $R_{ij}^2(\Phi)$ \footnote{We note that in Eq.~(1) of 
Ref.~\cite{Frodermann:2007ab} we inadvertently forgot the factor 2 on the 
left hand side of Eqs.~\eqref{eq:HBTosciamp}.}:
\begin{eqnarray}
  2 \frac{R^2_{(o,s),2}}{R^2_{s,0}} &=& 
    \frac{R^2_{(o,s)}(0) - R^2_{(o,s)}(\frac{\pi}{2})}
       {R^2_{s}(0)     + R^2_{s}(\frac{\pi}{2})}, 
\nonumber\\
  2 \frac{R^2_{os,2}}{R^2_{s,0}} &=& 
    \frac{R^2_{os}(\frac{\pi}{4}) - R^2_{os}(\frac{3\pi}{4})}
       {R^2_{s}(0)                + R^2_{s}(\frac{\pi}{2})}.
\label{eq:HBTosciamp}
\end{eqnarray}

The oscillation amplitudes for $R_o^2$ (Fig.~\ref{fg:photonAmps}a) and 
$R_s^2$ (Fig.~\ref{fg:photonAmps}b) differ mostly by an overall sign for 
both photons and pions; any additional differences in magnitude are due to 
azimuthal oscillations of the temporal contributions to $R_o^2$ 
\cite{Lisa:2000ip,Heinz:2002sq}. The latter are larger for pions than for 
photons. Fig.~\ref{fg:photonAmps}c shows that the azimuthal oscillations of 
the $R_{os}^2$ cross-term are much larger for photons than for pions (as 
already seen in Figs.~\ref{fg:photonosci1} and \ref{fg:pionHBTNC}), and that 
this difference is mostly due to the earlier emission of the photons (it is
more pronounced for the photons from the QGP phase than from the
hadron phase, but even in the HG phase this cross-term oscillates
more strongly for photons than for pions at large $K_\perp$ where
HG photons are emitted earlier than pions). 

The most important information can be extracted from the plot utilizes 
the oscillation amplitudes of $R_s^2$. Figure \ref{fg:photonAmps}b shows that 
$R_s^2$ oscillates more strongly for photons emitted from 
the QGP stage than from the hadron phase. This reflects the larger 
eccentricity of the fireball at earlier times. The oscillation amplitude for 
the total photon emission function interpolates between the QGP and HG limits: 
for $K_\perp\gtrsim2$\,GeV/$c$ it coincides with the QGP curve (indicating 
complete QGP dominance for $K_\perp>2$\,GeV/$c$), but even at $K_\perp=0$ the 
QGP contribution still plays a significant role. Retiere and 
Lisa \cite{Retiere:2003kf} showed for pions that one can extract the source 
eccentricity at freeze-out from the relation (see also Eq.~(2) in 
\cite{Heinz:2002sq}) 
\begin{equation}
\label{ecc}
  \epsilon_x = 2\,R^2_{s,2}/{R^2_{s,0}}
\end{equation}
in the limit $K_\perp\to 0$. In Ref.~\cite{Adams:2003ra} this procedure
was successfully applied to RHIC data. We have argued above that for 
photons a geometric interpretation of the r.h.s. in \eqref{ecc} should 
remain possible at non-zero $K_\perp$. If this is true, we can follow the 
fireball eccentricity backward in time by following the black solid 
line in the upper left panel of Fig.~\ref{fg:photonAmps} from low
to high $K_\perp$: $K_\perp=0$ would represent a time somewhere around 
$T_c$ whereas for $K_\perp > 2$\,GeV/$c$ we would probe times close
to thermalization of the QGP. Of course, fixed $K_\perp$ values cannot
be mapped one-to-one to sharply defined emission times -- the intention 
of our argument is to point out a qualitative and novel tendency that 
becomes accessible with photon interferometry. The analogous procedure
for pion HBT oscillations has no similar meaningful geometric 
interpretation.

\section{Conclusions}

In this paper we presented a first comprehensive study of two-photon
correlations and the azimuthal oscillations of the photon HBT radii
in non-central heavy-ion collisions. We based our investigation on
the hydrodynamic photon emission function for Au+Au collisions at 
$\sqrt{s}=200\,A$\,GeV with impact parameter $b=7$\,fm as a
model system. By comparing photon and pion HBT correlations we
were able to identify several key differences, both in the formalism
and in the numerical results, that make azimuthally sensitive photon 
interferometry an exciting prospect for future experimental studies.

The two most important insights from this study are (i) that since real 
photons are massless, a qualitative change of the shape of the 2-particle 
correlator $C(\bm{q})$ for soft photon pairs with $K \lesssim 1/
R_\mathrm{fireball}$ develops which, if it can be measured, will allow 
to separate the spatial and temporal aspects of photon emission in a 
model-independent fashion, and (ii) that at all transverse momenta within 
the range of validity of the hydrodynamic model photon emission is 
essentially volume dominated, with a strong QGP component from early 
times close to the fireball center and rather weak distortions from
a secondary flow-boosted hadronic emission component that extends 
more towards the edge of the fireball in the emission direction.
The first observation requires measuring 2-photon correlations
at very small pair momentum which will be very difficult; it also 
requires a careful shape analysis of the 2-photon correlator in
relative momentum since both the on-shell and smoothness approximations
break down in this $K_\perp$ range, and the correlator becomes strongly 
non-Gaussian. The second feature is more useful in practice since it works 
over a large range of pair momenta; it allows to use photon HBT radii for 
measuring the size and shape of the fireball at early times, before 
hadronization and hadronic freeze-out, and to map out the time evolution 
of its spatial eccentricity, by studying the normalized azimuthal oscillation
amplitude of the sideward radius as a function of transverse photon
pair momentum. This realizes a long-held dream for using 2-photon
correlations as a microscope for measuring the early fireball geometry. 
 
\section*{ACKNOWLEDGMENTS}
We gratefully acknowledge stimulating discussions with Mike Lisa who 
informed us that he, in unpublished work together with Giuseppe Verde, 
had previously seen evidence for breakdown of the on-shell and smoothness
approximations for pion pairs at very small transverse pair momentum. 
We also thank Joe Kapusta for helpful discussions and constructive comments
on an earlier version of this manuscript. Finally we thank Rupa Chatterjee and 
Dinesh Srivastava for some very infomative discussions. 
The work reported here is in part based 
on the PhD thesis of E.F. written at the Ohio State University. It was 
supported by the U.S. Department of Energy under grants DE-FG02-01ER41190 and 
DE-FG02-87ER40328.

\bibliography{photonHBT}

\end{document}